\documentclass[preprint2]{aastex}

\usepackage{natbib}
\usepackage{amsmath}
\usepackage{bm}
\usepackage{multirow}

\slugcomment{Accepted for publication in ApJ}

\shorttitle{The birth function of stellar populations}
\shortauthors{Gennaro et al.}

\begin{document}


\title{A new method for deriving the stellar birth function \\of resolved stellar populations\footnote{Based on observations made with the NASA/ESA \emph{Hubble Space Telescope}, obtained at STScI, which is operated by AURA, Inc., under NASA contract NAS 5-26555}}


\author{M.~Gennaro\altaffilmark{1},  K. Tchernyshyov\altaffilmark{2}, T.~M.~Brown\altaffilmark{1}, K. D. Gordon\altaffilmark{1,3}}
\email{gennaro@stsci.edu}
\affil{$^1$Space Telescope Science Institute, 3700 San Martin Drive, Baltimore, MD 21218}
\affil{$^2$Department of Physics and Astronomy, The Johns Hopkins University,\\ 3400 N. Charles Street, Baltimore, MD 21218}
\affil{$^3$Sterrenkundig Observatorium, Universiteit Gent,
 Gent, Belgium}



\begin{abstract}

We present a new method for deriving the stellar birth function (SBF) of resolved stellar populations. The SBF (stars born per unit mass, time, and metallicity) is the combination of the initial mass function (IMF), the star-formation history (SFH), and the metallicity distribution function (MDF). 
The framework of our analysis is that of Poisson Point Processes (PPPs), a class of statistical models suitable when dealing with points (stars) in a multidimensional space (the measurement space of multiple photometric bands). The theory of PPPs easily accommodates the modeling of measurement errors as well as that of incompleteness.
Compared to most of the tools used to study resolved stellar populations, our method avoids binning stars in the color-magnitude diagram and uses the entirety of the information (i.e., the whole likelihood function) for each data point; the proper combination of the individual likelihoods allows the computation of the posterior probability for the global population parameters. This includes unknowns such as the IMF slope and combination of SFH and MDF, which are rarely solved for simultaneously in the literature, however entangled and correlated they might be.
Our method also allows proper inclusion of nuisance parameters, such as distance and extinction distributions. The aim of this paper, is to assess the validity of this new approach under a range of assumptions, using only simulated data.  Forthcoming work will show applications to real data.
Although it has a broad scope of possible applications, we have developed this method to study multi-band HST observations of the Milky Way Bulge. Therefore we will focus on simulations with characteristics similar to those of the Galactic Bulge .

\end{abstract}



\section{Introduction}

The study of resolved stellar populations is going through a remarkable growth period, with space observatories like the Hubble Space Telescope (HST) providing high-resolution probes through nearby galaxies, and all-sky surveys like the Sloan Digital Sky Survey (SDSS) and Panoramic Survey Telescope and Rapid Response System (Pan-STARRS) revealing ever more substructure in the Local Group. Upcoming missions like the James Webb Space Telescope (JWST) and the Large Synoptic Survey Telescope (LSST) will further increase this wealth of data.  
A big challenge for the current and next generations of astronomers is that of developing appropriate tools to make the best out of these data, with sound statistical methods to help understand the underlying errors and properly interpret the results.
Ultimately, in the context of stellar populations, this means correctly interpreting the features observed in the color-magnitude diagram (CMD), or in equivalent diagrams.

The recent literature offers several examples of such methods that can be grouped into two broad categories: methods based on binning the data using a grid defined on the CMD, and methods that do not bin the data, but instead use each individual measurements separately.
The latter, at least in some respects, can be seen as the limit of the former in the case of very small grid cells where only 0 or 1 stars are observed in each cell.

Most of the bin-based methods, with their own differences, follow a similar approach. They first create a set of basis functions, i.e., simulated CMDs of simple stellar populations; these are usually realized with Monte Carlo techniques, accounting for the photometric errors and selection effects. They then linearly combine the basis functions to produce a synthetic CMD for the whole population. The fit is then performed via minimization of statistics comparing the predicted and observed numbers of stars in each grid-cell, in order to find the appropriate weight for each basis function. These weights correspond to the intensity of the star formation episode associated with each simple stellar population.
Examples of bin-based methods are those developed by \cite{Harris:2001fj}, \cite{Vergely:2002ve}, \cite{Dolphin:2002lr}, \cite{Ng:2002gf}, \cite{Cignoni:2006vn}, and \cite{Aparicio:2009uq}.
These methods have been successfully applied to a wide range of observations.
However, they become harder to apply when very few stars are observed, making a large number of CMD cells empty, or equivalently, forcing the use of very large cells. 
This limitation is mostly evident when more than 2 photometric bands data are available, making the number of useful cells even smaller, a problem that can be seen as one form of the curse of dimensionality.

Unbinned methods try and use the full information available for each datum, taking into account the noise associated with individual measurements.
Generally speaking, the methods in this class are based on computing the probability of each observed datum given the available stellar evolutionary models. The individual probabilities are then appropriately combined to derive the population parameters.
Again, several such methods are described in the literature, each with its own peculiarities
\citep[e.g.][]{Tolstoy:1996fk,Hernandez_1999,2005A&A...436..127J,2006MNRAS.373.1251N,Da-Rio:2010fk,Walmswell:2013yq}. The last paper also gives an insightful and detailed review of the differences between (and within) binned and unbinned methods.

To build our method we start by recognizing that, on a population scale, the star formation process can be described as a stochastic process in which individual stars are drawn independently from a parent distribution; this kind of process can be modeled as a type of random process known as a Poisson Point Process (PPP).
If we define our PPP on the space of intrinsic physical parameters, i.e. if we consider 
stellar age, mass, and metallicity as the stochastic variables, the parent distribution is simply the stellar birth function (SBF), which represents the probability of there being a star with a certain age, mass, and metallicity. The SBF can be regarded as a combination of the star formation history (SFH), initial mass function (IMF), and metallicity distribution function (MDF). 

The key to our method is that in the PPP formalism, one can map the probability distribution on the intrinsic parameter space to an equivalent probability distribution on a space of arbitrary observable quantities, such as (noisy) photometric observations or individual stellar spectra.
Additionally, this mapping can include the fact that some stars, while born and hence relevant to the SBF, cannot be observed; this part of the mapping is related to data incompleteness and is technically referred to as \emph{thinning}. For example, these stars may have evolved off the main sequence or be fainter than the detection limit of the specific observations.
The form of this mapping is such that for each observed star, we can easily compute the likelihood of any combination of intrinsic parameters, including nuisance parameters such as distance or reddening. 
These likelihoods are then combined to compute the posterior probability of any given SBF in a way that fully uses each individual measurement's information.

The statistical model we derive is similar to the one by \cite{2013ApJ...762..123W}, where the authors use a hierarchical Bayesian approach to obtain the posterior probability of the slope of the high-mass IMF in the context of stellar clusters analysis. Other than using a different path for reaching a similar model description, there are further differences between our approach and that by \cite{2013ApJ...762..123W}. For example, we are not being limited to studying a single parameter, since we include the whole SFH and MDF in addition to the IMF slope. Moreover, we simulated data to test our method, taking into account both noise and incompleteness
and we have developed and described a full numerical approach for the actual calculation of both the likelihood and completeness functions.
This is in contrast to the approach of  \cite{2013ApJ...762..123W} who do not derive the likelihoods from photometric measurements, but instead use an analytic approximation to describe the likelihoods of individual objects as well as the incompleteness function.
Within the astronomical community PPPs have been used to study problems in other areas; without the presumption of being exhaustive, but only to show the broad scope of application of PPPs, we mention \cite{2002MNRAS.335..151T,2011ApJ...742...38Y,2014ApJ...795...64F} in the area of exoplanet search and population census, \cite{2013A&A...559A..90L} who use PPPs to study the local Schmidt law in molecular clouds, and \cite{2007ASPC..371..417H} who model the Quasar Luminosity function in magnitude-redshift space as a PPP.

Our implementation of PPPs for SBF determination requires a discretization of the intrinsic physical parameter space on a grid. This is common to all the existing methods since all the mappings between physical parameters and observations are based on stellar models, which only exist for finite grids of parameters. Our method formally accounts for such approximation in its definition.
It is important to note that if the adopted grid of models is fine enough to resolve each star's likelihood function, the discretized approach is substantially equivalent to a completely continuous one.

The PPP formalism has several important advantages: (i) it is an exact and faithful mathematical analogue to the generally accepted idea of stellar population formation; (ii) it allows us to exploit synergies between subfields of astronomy and between astronomy and applied mathematics and statistics; (iii) it is conveniently modular; and (iv) it is flexible and extendible. 
Exactness (i) makes the formalism a useful and astrophysically motivated starting point for developing practical techniques. 
In particular, all of the existing CMD fitting techniques mentioned above can be derived from the fully general PPP formalism by taking various combinations of simplifying assumptions, approximations, and limits. 
Intra- and inter-disciplinary synergy (ii) simplifies the process of devising new computational techniques and verifying old ones.
For example, we use lessons learned from medical imaging, specifically positron emission tomography to find the best-fit SBF. 
Other investigators can now apply the vast statistical literature on different types of optimal approximate methods to vet existing methods (e.g. to show that they are theoretically unbiased and have good variance properties) and develop new ones. 
As we will show in the paper, the modularity of the SBF posterior probability is computationally convenient (iii), since it allows us to separately precompute several of the necessary quantities (e.g., the individual likelihoods) and makes the other computations involved parallelizable and therefore fast. 
This modularity also makes the formalism and method flexible and extendable (iv). 
For example, we use a specific set of stellar evolutionary libraries and photometric bands, but the formalism and method both apply to \emph{any} set of libraries and photometric bands. 
Moreover, the method can also be expanded to include other kinds of observations, such as spectra of individual stars, purely by modifying the likelihood and thinning functions. 
We will show examples in which spectroscopic constraints are incorporated to provide strong information on the populations' MDF.

In future work, we intend to exploit our method's flexibility to analyze the data from the Galactic Bulge Treasury Program \citep{2009AJ....137.3172B,2010ApJ...725L..19B}. This is a deep HST dataset which includes five photometric bands, is supplemented  by available ground-based stellar spectroscopy, and which targets four fields in the Galactic Bulge,  where there exists significant star-to-star distance and reddening variation.

The paper is structured as follows: we describe the basics of Poisson Point Processes in Section~\ref{sec:PPP}.
We then describe the adopted library of stellar models in Section~
\ref{sec:modgrid}. 
Section ~\ref{sec:NIL} deals with the treatment of measurement errors and incompleteness and how they affect the specifications of the individual likelihoods.
Section \ref{sec:MCMC} describes the explicit solution of the population properties.
We outline the test catalogs simulation process in Section~\ref{sec:simcat} and apply our method on such catalogs and show the results in Section~\ref{sec:res}.
We summarize our findings in Section~\ref{sec:summary}.

\section{Poisson Point Processes}
\label{sec:PPP}

A Poisson Point Process (PPP) is a statistical model that describes the counting of points in a multi-dimensional space.
A full description of PPPs and some of their applications is given in \cite{Streit}. Given the unfamiliar nature of these models to the astronomical community, in the following we summarize the basics of PPPs following the development in \cite{Streit}. In particular, we focus on the aspects of PPPs that are relevant to the analysis of stellar populations.

A realization of a PPP consists of a certain number of points observed in a state space $\mathcal{S}$. 
For our purposes, the state space is a 3D space in which the coordinates are stellar mass, age, and metallicity\footnote{Even if we denote the metallicity with the lowercase symbol $z$, we always make use of the spectroscopic definition of metallicity: [M/H] $ = \log \left( \frac{n_{\mathrm{metals}}}{n_\mathrm{H}} \right) - \log \left(\frac{n_{\mathrm{metals}}}{n_\mathrm{H}}\right)_{\odot} $. Because models can include variations in $\alpha$-element abundances, [Fe/H] and [M/H] are not necessarily the same. However, because a given set of models is computed using a well-defined [$\alpha$/Fe] vs.\ [Fe/H] relation, [Fe/H] will be sometimes used or mentioned instead of [M/H].}.
We will show in the following how the PPP of interest in the study of the SBF, which is defined in the stellar parameters space, can be related to a different PPP, i.e. the noisy incomplete set of photometric measurement we have access to.

One can easily imagine extending $\mathcal{S}$ to include other dimensions that can be of interest in the study of stellar populations, such as stellar distance, reddening, and multiplicity properties. The formalism would be equivalent and, for the sake of simplicity, we will drop these additional dimensions.
While the state space $\mathcal{S}$ can in principle be infinite, we are interested in realizations of PPPs on a bounded subset $\mathcal{R}$ of  $\mathcal{S}$.
For the purpose of stellar population analysis, the subset $\mathcal{R}$ will be defined by the range of stellar parameters in the specific set of adopted models (see Sect.~\ref{sec:modgrid}).

In describing a PPP, both the number and the distribution of points over the state space are random variables. 
A realization of a PPP in $\mathcal{R}$ is denoted by 
\begin{equation}
\xi = (n, \{ s_1,\cdots,s_n\})
\end{equation}
where the total number of points $n$ is explicitly indicated and the ordering of the $s_i$ points is irrelevant; the $s_i$ can, in principle, include duplicates.
The most important quantity that characterizes a PPP is its intensity, $\lambda (s)$, which describes how the $s_i$ points are distributed in the state space.  The intensity must be non-negative everywhere and has to integrate to a finite value over the state space:

\begin{equation}
\forall s \in \mathcal{S},\, \lambda(s) \geq 0 ;\quad 0 \leq \int_{\mathcal{R}} \lambda(s) \mathrm{d}s < \infty
\end{equation}

Two conditions must be met for a model to be a PPP: 
1) the total number of points in the subset $\mathcal{R}$, $N_{\mathcal{R}}$, has a Poisson distribution with parameter $\int_{\mathcal{R}} \lambda(s) \mathrm{d}s$ and 2) if $\mathcal{R}_1$ and $\mathcal{R}_2$ are disjoint, then $N_{\mathcal{R}_1}$ and $N_{\mathcal{R}_2}$ are independent.

Having set the stage, we show how a realization $\xi$ can be generated.
First, the number of points $n$ is a draw of the Poisson variable, $N$, distributed according to:
\begin{equation}
\label{eq:totnum}
p_N(n) = \frac{\mu^n}{n!} e^{-\mu}  \, ;\quad \mu \equiv \int_\mathcal{R} \lambda(s) \mathrm{d}s  \;.
\end{equation}
The integral of the intensity therefore gives the expected number of points.
The location of these points in $\mathcal{R}$ is given by $n$ independent draws of the random variable $S$ with probability distribution function (pdf) given by
\begin{equation}
\label{eq:pointdist}
p_{S}(s) = \frac{\lambda(s)}{\int_{\mathcal{R}} \lambda(s) \mathrm{d}s}
\end{equation}

We introduce a random variable $\Xi \equiv (N, \mathcal{X})$, where $N$ is the number of points and $\mathcal{X} =  \{s_1,\cdots,s_n\}$ is the points set.
The probability of a generic event evaluated at $\Xi = \xi$ is given by
\begin{equation}
p_{\Xi}(\xi) = p_N(n)\, p_{\mathcal{X}|N}(\,\{s_1,\cdots,s_n\}\,|\,n)
\end{equation}
where the first factor is given by Eq.~(\ref{eq:totnum}), the second is
\begin{equation}
 p_{\mathcal{X}|N}(\,\{s_1,\cdots,s_n\}\,|\,n) = n! \prod_{i=1}^n p_{S}(s_i),
\end{equation}
and $p_{S}$ is the pdf for a single point, as given by Eq.~(\ref{eq:pointdist}). The $n!$ factor is due to the fact that there are $n!$ possible combinations of the $s_i$'s that correspond to the unordered set $\mathcal{X}$.
Combining everything we obtain that 
\begin{equation}
\label{eq:PPPdef1}
p_{\Xi}(\xi) = e^{-\mu} \prod_{i=1}^n \lambda(s_i)
\end{equation}

In the context of stellar populations, the intensity function corresponds to the stellar birth function, i.e., represents how many stars have been formed per unit time, mass, and metallicity, within the patch of the sky under study. 
This function is a combination of the initial mass function (IMF), the star-formation history (SFH), and the metallicity distribution function (MDF).
While in principle we might expect the IMF to be a function of both age and metallicity, in the following we will make the simplifying assumption that it is instead independent with respect to both. 
Therefore $\lambda(s) = \lambda(m,a,z) \propto \mathcal{I}(m)\,\Phi(a,z)$.

Equation~(\ref{eq:PPPdef1}) gives the probability of $n$ stars being formed with a given set of properties $\{(m,a,z)_1, \cdots, (m,a,z)_n\}$, i.e., it is the probability of the set of physical parameters $(m,a,z)$, given $\lambda$; $p_{\Xi}(\xi) = p_{\Xi}(\xi\, |\,\lambda)$.
Using Bayes' theorem, we can write the probability of $\lambda$ given the available $\{(m,a,z)\}$:
\begin{equation}
\label{eq:BayPPP}
p_{\Lambda}(\lambda\, |\,\xi) \propto p_{\Xi}(\xi\, |\,\lambda)\, p_{\Lambda} (\lambda)
\end{equation}
where the normalization factor is omitted.
If we had the parameters $(m,a,z)$ for all the stars in our sample, our computational problem would be exploring the pdf of $\lambda$ defined in Eq.~(\ref{eq:BayPPP}).
However, the problem to solve in the case of stellar populations is that of determining $\lambda$ given the number of observed stars and a set of flux or magnitude measurements for each star. 
The physical parameters $(m,a,z)$ are not directly observable or accessible, so stellar models must be used to interpret the measured quantities in terms of $(m,a,z)$. The uncertainties in these physical parameters will depend upon the measurement and modeling errors.
Furthermore, of the stars that are born according to the true, underlying $\lambda$, only some will be observable. The incompleteness of the data is due to both stellar evolution (massive stars ending their lives as stellar remnants) and observational limits (the ability to detect an object and of measure its flux given the noise and crowding properties of the specific observations).

The theory of PPPs, as outlined in \cite{Streit}, easily accommodates measurement processes and incompleteness\footnote{Incompleteness is technically referred to as \emph{thinning}.}. We summarize the treatment of both below.

\subsection{Measurement process and errors}
The unknown intensity $\lambda$ assumes values in the space $\mathcal{S}$ of stellar mass, age and metallicity. The actual measurements are made in another space, which we indicate with $\mathcal{T}$ and which can be thought of as a $k$-dimensional magnitude space, where $k$ is the number of available bands.
\footnote{It may be argued that for all modern imaging systems, the fundamental observed quantities are fluxes, or better, photon counts (or count rates for near infrared detectors). Magnitudes are derived quantities and, given the non linear relation between fluxes and magnitudes, the noise characteristics are definitely different between them. However, in our ideal experiments we assume that we have complete knowledge of the noise in magnitude space. Likewise, in realistic applications, the noise can be estimated in either flux or magnitude space using artificial stars experiments. As long as the noise is treated consistently, it shouldn't matter which variable is considered.}

For a set of stellar parameters $s=(m,a,z)$, we can determine the probability $p(\,t\vert\,s)$ of observing a set of magnitudes $t=(M_1,\cdots, M_k)$ using stellar models and observational uncertainties.
Given a realization 
$\xi = (n, \{ (m,a,z)_1, \cdots, (m,a,z)_n \} )$ of a PPP with intensity $\lambda$, it can be shown that $\eta = (n, \{ (M_1,\cdots,M_k)_1,\cdots,(M_1,\cdots,M_k)_n\})$ is a realization of a PPP with intensity equal to:
\begin{equation}
\label{eq:PPPmeasur}
\nu(t) = \int_{\mathcal{R}} p(\,t\,|\,s) \, \lambda(s) \,\mathrm{d}s \;.
\end{equation}

Computing the likelihood $p(\,t\,|\,s\,)$ requires knowledge of the underlying noise properties. We will show in section \ref{sec:NIL} how we determine $p(\,t\,|\,s\,)$ for our simulated catalogs.

\subsection{Incompleteness}
\label{sec:incompl}
When a stellar field is observed, some of the stars that actually formed within that field cannot be detected. They may have evolved into stellar remnants (white dwarfs, neutron stars and black holes) or been completely destroyed by deflagration. Some of the remnants may technically be observable, but they do not end up in regions of the CMD \footnote{Even if we refer to CMDs and display relevant figures using this tool, we emphasize that our method treats magnitudes as the real variables, as only individual magnitudes are the product of the measurement process, while colors are derived quantities.} that can be studied using stellar evolution models in the traditional sense. We currently neglect all evolution beyond the red giant branch (RGB) phase in our treatment. The rapid post-RGB evolutionary phases could be re-introduced into our framework, provided the necessary models for the later stages of stellar evolution are included.

Other stars might instead not be observable because they are intrinsically too faint (below the detection threshold) or are at very small angular separation from brighter neighbors that hamper their detection (an effect known as crowding or confusion).
Whatever the source of incompleteness, it must be accurately modeled and can be included in the framework of PPPs.
If we indicate with $0\leq \alpha(m,a,z) \leq 1$ the probability of detecting a star with a given mass, age and metallicity, given a realization $\xi = (n, \{ (m,a,z)_1, \cdots, (m,a,z)_n \} )$, the corresponding incomplete realization is obtained by retaining each $(m,a,z)_i$ with probability $\alpha( (m,a,z)_i)$.
The incomplete realization is given by $\xi_{\alpha} = (l, \{ (m,a,z)_1, \cdots, (m,a,z)_l \} )$ with $l\leq n$.
It is possible to show that the incomplete process is still a PPP, with intensity:
\begin{equation}
\label{eq:thinning}
\lambda_{\alpha}(m,a,z) = \lambda(m,a,z)\,\alpha(m,a,z) 
\end{equation}

It is important to notice that incompleteness is a property of the models and is a fundamental part of the model definition. In section \ref{sec:NIL} we will show a possible way to estimate $\alpha(m,a,z)$.

\subsection{Recap: PPPs for noisy, incomplete photometric data and a discrete parameter space}
\label{sec:summLambda}

Combining everything together, the measurements that one has after observing a patch of the sky and performing photometry on the resulting images is a realization of an incomplete, noisy PPP in a $k$-dimensional magnitude space. We are interested in going from these measurements to a solution for $\lambda$, a complete PPP in the 3-dimensional space of physical parameters $(m,a,z)$.
Solving for $\lambda$ means solving for a continuous function over the whole $\mathcal{R}$ space. This can be accomplished by making some simplifying assumptions.
First, we assume that the IMF is a power-law with a single slope $\gamma$, $\mathcal{I}(m)=\frac{\mathrm{d}n}{\mathrm{d}m}\propto m^{\gamma}$. 
Second, we assume that $\Phi(a,z)$, the combined SFH-MDF, is piecewise-constant. In our implementation, the constant intervals are evenly spaced; their centers form a regular grid. This simplification is equivalent to discretizing the age-metallicity parameter space and weighting each grid point by the implied grid-cell's volume.
Solving for $\lambda$ is equivalent to solving for the slope $\gamma$ and for the number of stars that have formed within each $(a,z)$ cell:
$p(\,\lambda(m,a,z)\,) \equiv p(\,\gamma,\{n_{i_{a,z}}\} \,)$; for ease of notation, $i_{a,z}$ or $i_{m,a,z}$ indicate tuples of indices, $(i_a,i_z)$ and $(i_m,i_a,i_z)$, respectively.

The effective discretization of the parameter space implies a substitution of the integral in Eq. (\ref{eq:PPPmeasur}) with a sum over the cells.
The final probability for $\lambda$, given the incomplete measurements $\eta = (l, \{ (M_1,\cdots,M_k)_1,\cdots,(M_1,\cdots,M_k)_l\})$ is thus

\begin{equation}
\label{eq:finPPP}
p( \gamma,\{n_{i_{a,z}}\} \, | \, \eta)  \propto   e^{ - \mu_{\alpha} } \left( \prod_l \nu_{\alpha,l} \right)\; p(\gamma,\{n_{i_{a,z}}\}),
\end{equation}
where
\begin{equation}
\label{eq:norm}
\mu_{\alpha} = \sum_{i_{m,a,z}} n_{i_{a,z}} w_{i_m} \alpha_{i_{m,a,z}}.
\end{equation}
The indices ${i_{m,a,z}} = (i_m,\,i_a,\,i_z)$ run over the mass, age and metallicity cells, and $w_{i_m}$ is the integral of the IMF across the $i_m$-th cell, which has a simple analytic expression in the case of a single power law. The index $l$ runs over the observed stars.
Because the IMF is normalized, we have that $\sum_{i_{m}} n_{i_{a,z}} w_{i_m} = n_{i_{a,z}} \sum_{i_{m}}  w_{i_m} = 
n_{i_{a,z}}$.

The $\nu_{\alpha,l}$ are integrals of incomplete measurement process intensity over the parameter space (see Eq.~\ref{eq:PPPmeasur} and Sect.~\ref{sec:incompl}). Explicitly,

\begin{equation}
\label{eq:indlkl}
\nu_{\alpha,l} = \sum_{i_{m,a,z}} n_{i_{a,z}} w_{i_m} \alpha_{i_{m,a,z}} \,f_{i_{m,a,z}}(l),
\end{equation}
where
\begin{multline}
f_{i_{m,a,z}}(l) = \\
 \int_{\mathcal{R}_{i_{m,a,z}}} p(\,(M_1,\cdots,M_k)_l \,|\, m, a, z)\, \mathrm{d}m \,\mathrm{d}a\,\mathrm{d}z
\end{multline}
is the integral of the $l$-th star's likelihood function over the $\mathcal{R}-\text{cell}$ corresponding to $(i_m, i_a, i_z)$; we show how $f(l)$ is calculated in Sect.~\ref{sec:indlik}.

To complete the model specification, we need to choose a prior $p(\gamma,\{n_{i_{a,z}}\})$. We have already specified part of the prior by choosing the range covered by the (m,a,z) grid. We discuss the rest of the prior specification in Sect.~\ref{sec:MCMC}.

Equations (\ref{eq:norm}) and (\ref{eq:indlkl}) can be modified to add further dimensions to the problem, e.g., nuisance parameters such as distance or extinction, indicated globally by $\pi$. 
Adding additional parameters requires defining the incompleteness, stellar likelihoods, and intensity prior over the expanded parameter set $(m, a, z, \pi)$. 
If we do not attempt to solve for the intensity as a function of $\pi$, we can marginalize over $\pi$ before attempting to solve for $\lambda(m,a,z)$;
this alters the incompleteness, stellar likelihoods, and intensity prior by a (possibly different) constant multiplicative factor at each $(m,a,z)$ value.
We will show an application involving nuisance parameters in Sect.\ref{sec:nuipar}.

We will show in Sect.~\ref{sec:NIL} how the individual terms of Eq.~\ref{eq:finPPP} can be computed.
The practical solution of the problem of computing $p( \gamma,\{n_{i_{a,z}}\} \, | \, \eta)$ will be given in Sect.~\ref{sec:MCMC}, where we apply a Markov Chain Monte Carlo (MCMC) algorithm to generate samples from this distribution.

One of the fundamental assumptions underlying our model is that the stars are independently independently drawn from the IMF.
Such an assumption might not necessarily hold true, according to some theories of star formation. In that case our method (and all the methods that assume that there is an IMF) would fail.
The impact of such assumption might be more severe in the study of young massive clusters, where feedback is particularly important.
However we believe that for the study of the Galactic Bulge the independence condition is satisfied. In the Bulge or in other regions not actively forming stars, stars that are observed within one patch of the sky, have formed possibly in different regions, at different times, and have undergone dynamical mixing. Therefore, at their formation, they were truly independent. If this is the case, we can still try and infer the slope of an IMF that can be thought as a parent distribution averaged across the formation history of that population.

\section{The model grid}
\label{sec:modgrid}

In order to convert the measured magnitudes into physical parameters, i.e., in order to interpret the data in terms of meaningful quantities, it is necessary to adopt stellar evolutionary models. For our examples, we use 
models computed with the Victoria-Regina code \citep{2012ApJ...755...15V}, updated with a  heavy element mixture suited for the stellar populations of the Galactic Bulge \citep{2014ApJ...794...72V}. 
The transformations from the model physical parameters $(\log L/L_{\odot}, \log T_{\mathrm{eff}}, \log g)$ to the observable magnitudes are performed using synthetic spectra computed with the MARCS stellar model atmospheres code \citep{2008A&A...486..951G}.
As explained in Sect.~\ref{sec:NIL}, we adopt one specific filter system, the one for the Wide Field Camera 3 (WFC3) onboard the {\it Hubble Space Telescope} ({\it HST}); however the scope of our method is not limited to one set of stellar models or a particular suite of photometric bands. Assessment of the systematic uncertainties related to the use of different stellar models, as well as the exploration of the best possible combination of filters for deriving IMF, SFH and MDF of resolved stellar populations is beyond the scope of this paper.

For this work, we use a grid with a 2\% spacing in mass, 2.5\% in age, and 0.1 dex in [Fe/H].
The mass and age steps correspond to a constant spacing in the logarithm of mass and age, respectively.
This type of spacing translates into a conveniently more constant spacing between stellar models in the CMD.
However, the grid spacing does not have to be regular for our technique to be applicable. 
Given that this method was developed to deal with multi-band photometric data for the Galactic bulge, we choose a range of parameters that is suitable for the bulge stellar population; again, the grid range can be changed and adapted to different problems. Specifically, we have $m \in [0.2,1.5]\; M_{\odot}$, $a \in [7.,14.7]$ Gyr, $z \in [-2.0,+0.4]$ dex.

The choice of the grid size and grid resolution has a practical impact on the method and is a matter of compromise between computing time and the ability to resolve rapid changes in $\Phi(a,z)$. 
The number of unknowns in our problem is almost equal to the number of age-metallicity cells. Because we choose to parametrize the IMF as a single power law,  one additional parameter is the slope of the IMF, $\gamma$. More complex models for the IMF could in principle be chosen in other applications, if there is reason to  think that --or to explore whether-- the data may help constrain an IMF model with increased complexity. 

The choice of the mass, age, and metallicity grid resolution must also be guided by the quality of the data at hand. 
One should be able to resolve the individual likelihoods without computationally expensive excessive oversampling. Unfortunately, stars in different evolutionary phases have very different likelihoods, in terms of how diffuse each likelihood is in the physical parameter space $(m,a,z)$. 
For example, if we can resolve a turnoff star's age likelihood function, we are oversampling the main sequence stars' age likelihood functions.
We can use the typical photometric errors in the turnoff region to estimate the precision attainable in age determination and then use a fraction of that as the age-step of the grid. The use of metallicity-sensitive photometric bands can increase the resolving power of our observations for [Fe/H]. 
The WFC3/UVIS $F390W$ band, for example, covers a spectral region that, in dwarf stars, contains strong metal lines.

\section{Noise, incompleteness and likelihoods.}
\label{sec:NIL}

We test the effectiveness of our method by simulating data and comparing the input and recovered intensity functions.
The simulated quantities are stellar magnitudes.
We use Gaussian noise for the simulated magnitude measurements and assume that the only errors involved are random errors. 
We further assume that the noise in multiple photometric bands for the same star is uncorrelated. For catalog generation and incompleteness evaluation, we also assume that the detections in each band are independent of each other, and that a star is considered detected when it is detected independently in all bands.

To estimate realistic noise levels for a typical stellar population study, we use data from \cite{2009AJ....137.3172B,2010ApJ...725L..19B}. 
These papers present observations of four stellar fields towards the Galactic bulge. 
The data from the full observing program will be analyzed in a forthcoming paper; for testing our method, we use the subset of the data that was taken in the OGLE29 low-extinction window. We also limit ourselves to 3 of the 5 bands, namely F390W, F555W, and F814W, respectively approximating Washington $C$, Johnson $V$, and Johnson $I$, which is sufficient to demonstrate our method. 
Because our simulations do not include distance and extinction as parameters, we convert the apparent magnitudes into absolute magnitudes by subtracting the average bulge distance modulus and typical extinction in each of the 3 bands.

\begin{figure*}[!t]
\centering
\includegraphics[width=\textwidth]{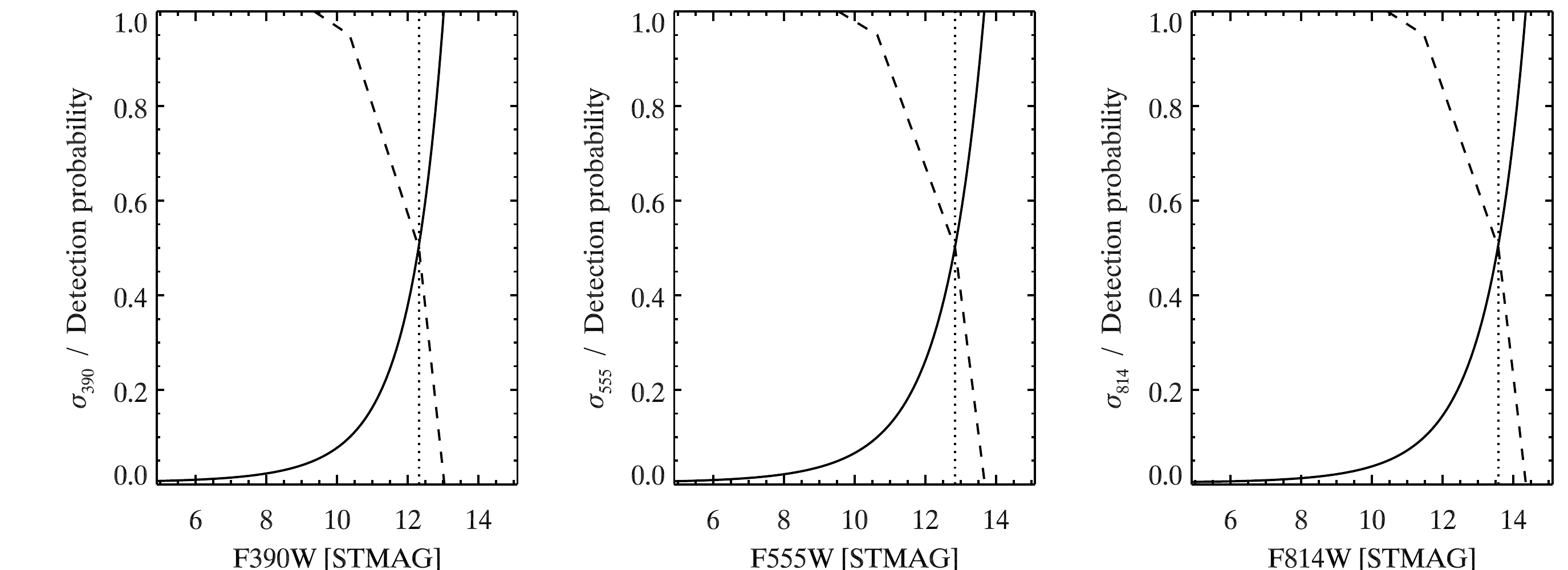}
\caption{Realistic values of the typical error and completeness derived from data by \cite{2009AJ....137.3172B,2010ApJ...725L..19B}, for the three photometric bands considered in this work. The x-axis indicate the measured magnitudes. Solid lines: error curves; dashed lines: completeness (detection probability); dotted vertical lines: 50\% completeness limit for that band.}
\label{fig:errinc}
\end{figure*}

For each band, we use the real data to estimate a relation between the observed magnitudes and photometric uncertainties.
We first bin the observed stars in magnitude bins and take the average photometric uncertainty in each bin. The uncertainties come from PSF fitting photometry using \texttt{DAOPHOT} \citep{1987PASP...99..191S}.
We then fit a third-order polynomial to the logarithm of the average uncertainty as a function of bin magnitude.
The fit coefficients are then used to compute the typical photometric error as a function of magnitude (see Fig.~\ref{fig:errinc}). 
 
We also adopt an incompleteness vs. magnitude relation. This relation is obtained by imposing that objects with magnitude errors of $0.05, 0.1, 0.5,\text{ and }1$ mag are detected in 100\%, 95\%, 50\%, and 0\% of the cases, respectively, and linearly interpolating the incompleteness between these magnitude values.
These incompleteness curves are shown in Fig.~\ref{fig:errinc}.
As in the data of \cite{2009AJ....137.3172B,2010ApJ...725L..19B}, the F555W and F814W bands have lower uncertainties and incompletenesses at fixed magnitude than the F390W band.
This means that the overall incompleteness of the simulated observations is generally dominated by F390W.
There could be situations in which completeness does not only depend on the stellar magnitudes, but also on the position across the field of view. This is specially true for dense stellar clusters, given the gradient in the crowding properties between the center and outskirts \citep[see e.g.][]{2011MNRAS.412.2469G}.
However the present work has been developed to deal with HST observations of the Galactic Bulge \citep{2009AJ....137.3172B,2010ApJ...725L..19B}, with a small field of view ($\sim 4$ arcmin) across which crowding is very homogeneous. 
We will limit ourselves to this simple situation where incompleteness does not depend on position.

\subsection{Computing $\alpha(m,a,z)$}
\label{sec:compcomp}
The incompleteness curves of Fig.~\ref{fig:errinc} represent the probability of detecting a star given its {\it measured} magnitude. 
The function $\alpha(m,a,z)$ represents the probability of detecting a {\it model} star, for which we know the intrinsic, or error-free, magnitudes $(M_{390},M_{555},M_{814})^{\mathrm{intr}}$.
At a single value of $(m,a,z)$, the model incompleteness is an average of the measured-magnitude incompleteness over the measured-magnitude pdf. 
Since we require a star to be detected in all bands, the incompleteness of a vector of measurements $(M_{390},M_{555},M_{814})^{\mathrm{obs}}$ is the product of the incompleteness in each band.

To compute $\alpha$ for each model grid cell centered on $(m,a,z)$, we simulate the attempted measurement of $j=1,\cdots,1000$ stars per grid cell. The 1000 $(m,a,z)_j^{\mathrm{intr}}$ values are randomly uniformly extracted within the cell. For each of them, $(M_{390},M_{555},M_{814})_j^{\mathrm{intr}}$ is computed using the library of stellar models. Corresponding $(M_{390},M_{555},M_{814})_j^{\mathrm{obs}}$ are extracted from Gaussians centered on $(M_{390},M_{555},M_{814})_j^{\mathrm{intr}}$ with the appropriate $\sigma$'s. Finally, using the curves in Fig.~\ref{fig:errinc} we decide whether that star would have been observed or not. This is done by comparing, for each band, the value of the detection probability with a uniform random number between 0 and 1. If a star is not detected in one band, then it is considered not detected at all.
The number of recovered stars divided by 1000 is the approximate value of $\alpha(m,a,z)$ averaged over the model cell.

There are several simplifying assumptions in our treatment. In general, for real observations, the explicit form of the noise is not known (even though it is often assumed to be Gaussian).
However, in the case of real observations, where real images of stellar fields are used,  the process described above can be reproduced using artificial star tests.
The latter consist of introducing into the images under study stars of know magnitudes (directly related to known $(m,a,z)$ through stellar models)  and then trying to measure the output magnitudes.
By using the exact same algorithmic sequence (including cuts, a posteriori selections, detection criteria and so on) as used for building the real catalog of observations, we can prepare a list of input vs. output (detected and undetected) stars.
For the computation of $\alpha(m,a,z)$ it is still possible to generate $(m,a,z)_j^{\mathrm{intr}}$ and the corresponding $(M_{390},M_{555},M_{814})_j^{\mathrm{intr}}$ from stellar models.
The list of input stars can be searched for stars with these magnitudes and the corresponding output can be checked to see whether it corresponds to a valid detection.
Finally, the ratio of detection-to-input stars can be used as $\alpha(m,a,z)$.

The above description deals with incompleteness due to the measurement process and crowding. These phenomena mostly affect models towards the faint (low-mass) end of the parameters grid. However, for the treatment of PPPs, we also need to consider stars that contribute to $\lambda$ but are no longer observable because they have evolved away from the main sequence.
These stars are instead towards the high-mass end of the model grid. 
To account for these lost stars, it suffices to check whether the $(m,a,z)_j^{\mathrm{intr}}$ points that are generated per grid cell in the previous step correspond to ``alive'' stars.
If they do not then they are simply considered as non-detected models and contribute to lowering $\alpha$ for that cell.

For real observations there might be cases where further \emph{a posteriori} cuts are imposed to photometric catalogs. 
For example, stars near the detection threshold are often removed because their measurement uncertainties are difficult to estimate.

In our simplified scheme, we assume we know everything about the noise model and we also mention that, in realistic cases, artificial star tests can be used to explore the noise characteristics and derive the incompleteness function.
However, even artificial star experiments may not capture all possible sources of errors and, in general, one might be wary of trusting the noisiest detections. 
If this is the case, any hard magnitude cut imposed to a catalog must be included in the computation of $\alpha(m,a,z)$. The only difference with the previously illustrated scheme is that the simulated values $(M_{390},M_{555},M_{814})_j^{\mathrm{obs}}$ must also fall above the imposed thresholds in order to be accounted for as detections. 
We will show the impact of such cuts on the recovery of $\lambda$ in Sect.~\ref{sec:res}.

\begin{figure*}[!t]
\centering
\includegraphics[width=.97\textwidth]{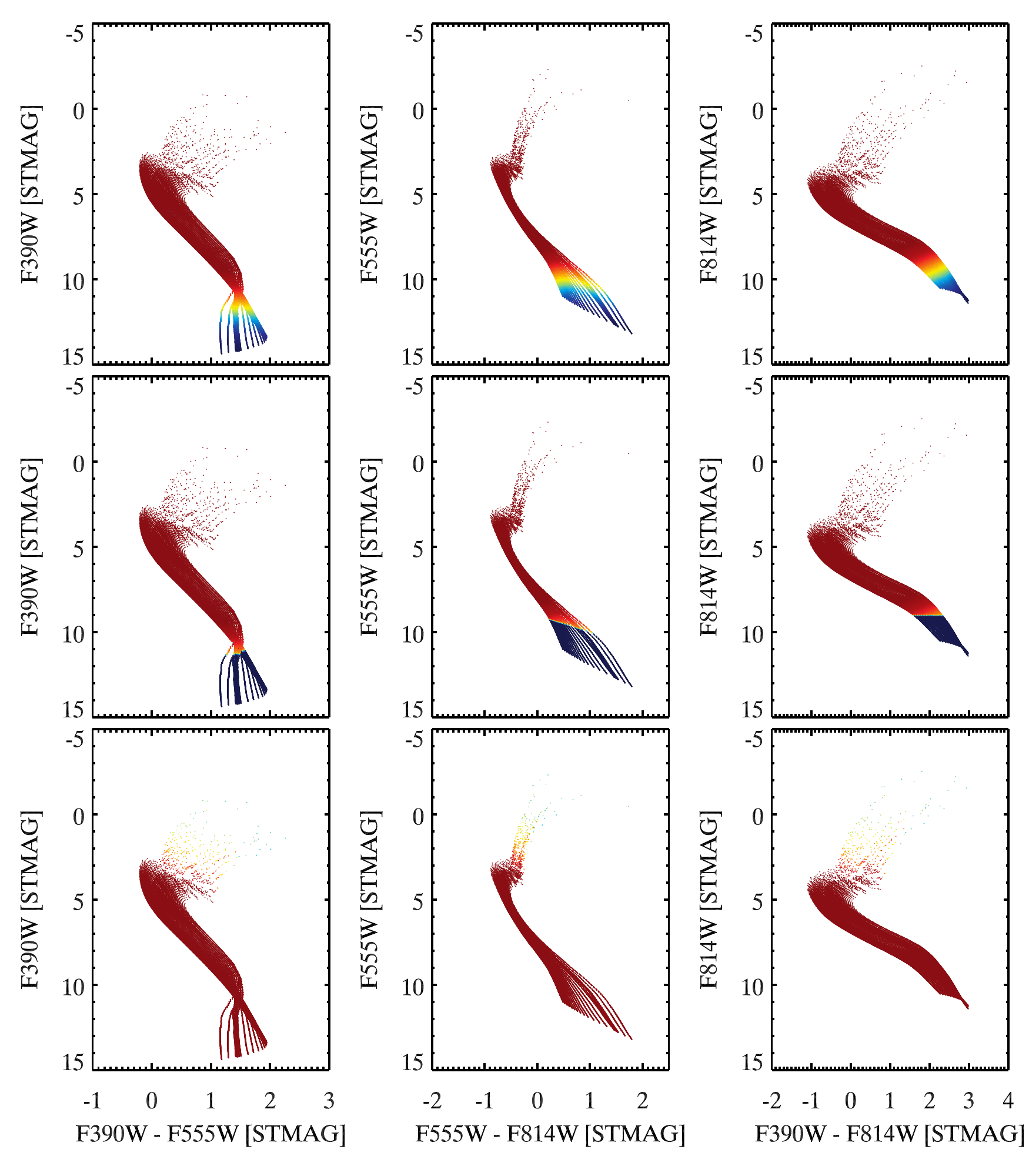}
\caption{CMDs of our model grids, color coded by incompleteness values. Red corresponds to 100\% complete models, blue to 0\%. The first two rows show the effect of incompleteness due to missing detections when no \emph{a posteriori} cut is applied to the photometric catalogs (top row) or when a cut at $M_{814}=9$ mag is applied (center row). The bottom row shows the contribution to incompleteness from stellar evolution. Red corresponds to model cells that are not evolved away from the main sequence; bluer colors correspond to cells where some fraction of the models is not observable, having evolved away from the main sequence.
}
\label{fig:comp9}
\end{figure*}

Figure \ref{fig:comp9} shows $\alpha(m,a,z)$ for our grid of models assuming the standard curves of Fig.~\ref{fig:errinc}; incompleteness is color-coded from dark blue ( $\alpha(m,a,z) = 0$) to red ( $\alpha(m,a,z) =1$) . 
Each model point corresponds to the center of a $(m,a,z)$ cell.
The CMDs in the top row show the values of $\alpha(m,a,z)$ when no cut is applied, while in the middle row, a cut at $M_{814}=9$ mag was applied, hence the sharp drop in incompleteness. To illustrate the effects of stellar evolution on incompleteness we separate the contribution to $\alpha$ from stars that have evolved, shown in the bottom row; here it is possible to notice that for the most massive model cells (in the red giant phase) a fraction of the cell volume corresponds to stars that are no longer observable, even when the central values $(m,a,z)$ for that cell correspond to still-alive stars.
For our method, the total $\alpha(m,a,z)$ is the product of the incompleteness in the first (or second) row with the incompleteness in the third row.

\subsection{Individual stellar likelihoods}
\label{sec:indlik}

\begin{figure*}[!t]
\centering
\includegraphics[width=.48\textwidth]{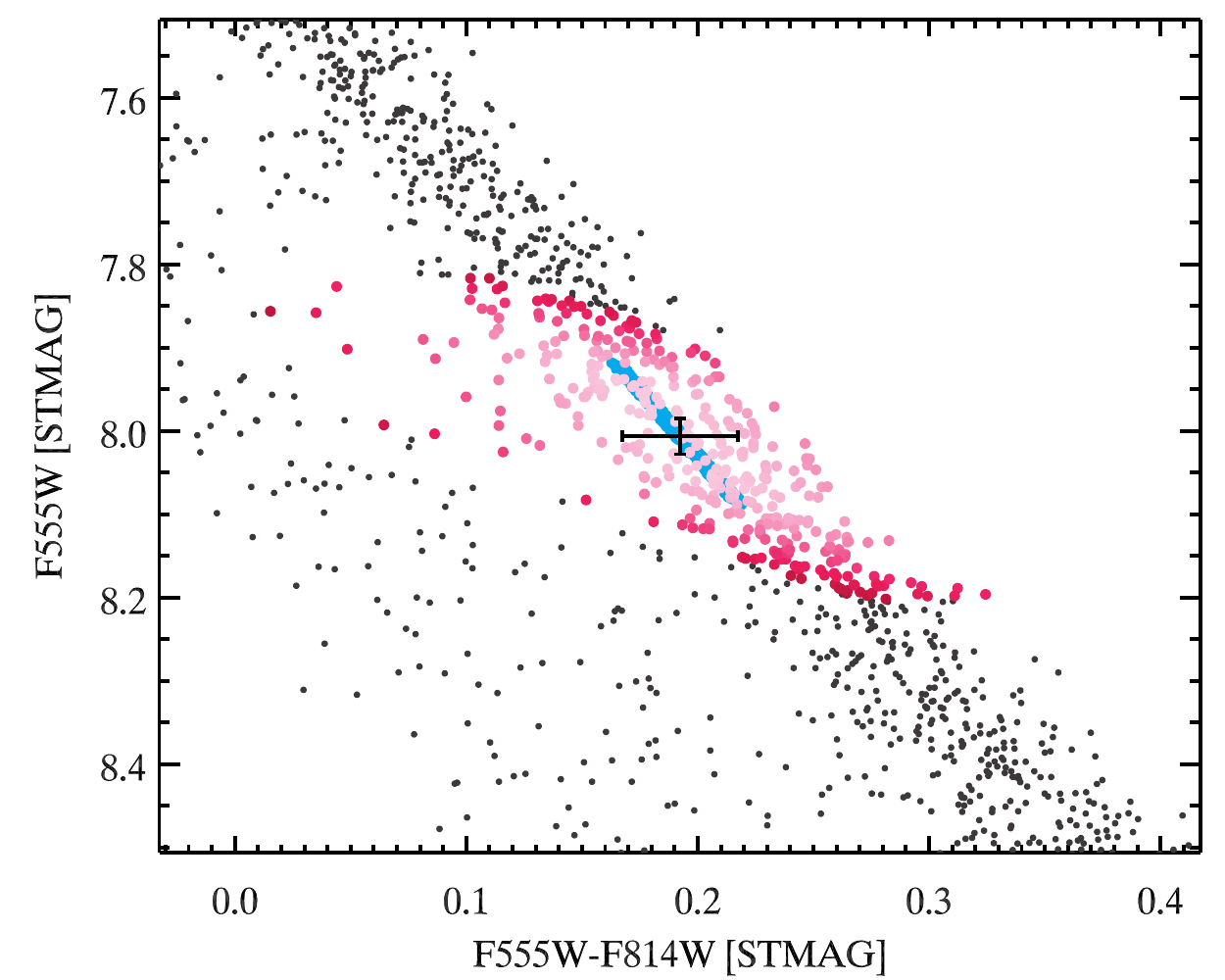}
\includegraphics[width=.48\textwidth]{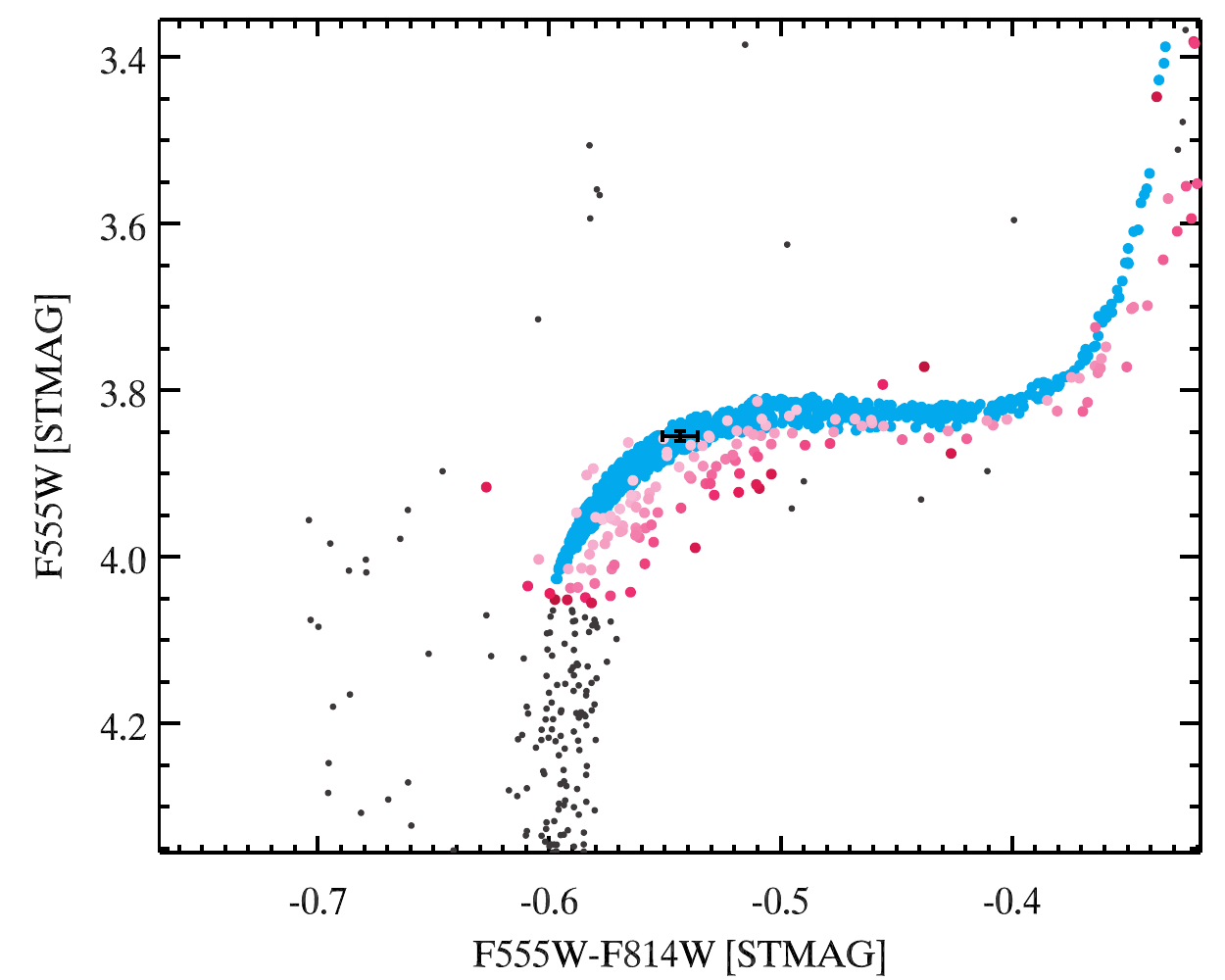}
\caption{Illustration of our likelihood calculation method based on averages over the cells. The blue points represent 500 model realizations within cells centered on $M=0.59 M_{\odot}\,\mbox{ (left), and}\, 0.92 M_{\odot}$, (right). Both model cells are centered on $a=12.05$ Gyr and [Fe/H]~$ = -0.3$ dex. The black points are simulated measurements from one of our artificial catalogs ({\sc Blg}, see Sect.~\ref{sec:simcat}). The pink points are data that are within $5\sigma$ from at least one of the 500 realizations (where $\sigma$ is the uncertainty associated to each model realization, according to the curves of Fig.~\ref{fig:errinc}). The shade goes from high (light pink) to low likelihood (dark pink). The error symbols correspond to the models with parameters in the center of the grid cells, and their size is equal to the typical $\sigma$ of that cell.
}
\label{fig:lklex}
\end{figure*}

An individual likelihood is the probability of detecting a star at $(M_{390},M_{555},M_{814})^{\mathrm{obs}}$ given the model parameters $(m,a,z)$.
The discretization of Eq.~(\ref{eq:indlkl}) implies that the likelihoods, $f_{i_{m,a,z}}(l)$ need to be integrated across the individual cells of the parameters grid.
To compute $f(l)$, we use the following method.
For each parameter grid cell we generate 500 values of $(m,a,z)_j$ within the cell. We compute the corresponding magnitudes $(M_{390},M_{555},M_{814})_j^{\mathrm{mod}}$ using the stellar models library. We then evaluate the photometric $\sigma$'s at those magnitudes. 
The magnitudes and $\sigma$'s completely characterize the Gaussian likelihoods; for each star $i$ in a simulated catalog, with magnitudes $(M_{390},M_{555},M_{814})_i^{\mathrm{obs}}$ we compute the value of 
\begin{multline}
p( (M_{390},M_{555},M_{814})_i^{\mathrm{obs}}\,|\,(m,a,z)_j) = \\
\left(\prod_{\mathrm{filt}} \frac{1}{\sqrt{2\pi}\sigma_{\mathrm{filt},j}}\right) \times \exp( -\chi^2/2)
\end{multline}

with
\begin{equation}
\chi^2 = \sum_{\substack{\mathrm{filt} = \\ 390,555,814}} {\left( \frac{ M^{mod}_{\mathrm{filt,j}}-M^{obs}_{\mathrm{filt,i}}}{\sigma_{\mathrm{filt},j}} \right)}^2.
\end{equation}

We then average the likelihood over the 500 $j$-values. This corresponds to a marginalization of the likelihoods across the model grid cell. 

The necessity of averaging the likelihoods over a model cell, instead of just taking the likelihood value for, e.g., the cell center, is apparent when considering evolutionary phases in which a star's position on the CMD rapidly changes.
While on the main sequence the changes are very slow, once a star reaches the turnoff region it may move a substantial length (in both magnitude and color) within a very short time.
Because these fast phases happen at different times for different stellar masses and metallicities, it would be computationally very intensive to define the models on a grid that is fine enough to resolve the fastest stellar evolutionary phases for all $(m,a,z)$ combinations.
However, the average across a cell allows for a proper treatment of these phases even within the limits of the grid resolution.
Figure~\ref{fig:lklex} shows an example of such a situation for one of our simulated catalogs (black points), focusing on two different regions of the CMD: the main sequence on the left and the turnoff on the right. 
In light-blue we show the 500 models realizations for two different cells, centered on $M=0.59 \,\mbox{and}\, 0.92 M_{\odot}$ respectively. Both cells are centered on $a=12.05$ Gyr and [Fe/H]~$ = -0.3$ dex. The width of the cells is 2\% in mass, 2.5\% in age and 0.1 dex in metallicity.
The light-blue area is the transformation of the physical parameter cube $(m\pm\delta m, a\pm\delta a, z\pm\delta z)$ into the CMD.

The models at the cell centers are identified by the error symbols, with error bar size equal to the average $\sigma$ on the cell.
The symbols in pink are simulated catalog stars that are within $5\sigma$ of at least one of the 500 realizations (where $\sigma$  is the uncertainty associated to each model realization, according to the curves of Fig.~\ref{fig:errinc}). The shades of pink correspond to high-to-low (light-to-dark) likelihood values on a logarithmic scale.
The need to compute likelihoods over entire model cells is particularly clear in the right panel. 
If we had considered only the model corresponding to the grid cell center, many data points would have been too far (in units of $\sigma$) to have any likelihood at that cell even though their physical properties are very close to the physical properties of the cell's center.
Averaging the likelihood over the entire model cell helps prevent this problem.

\begin{figure*}[!t]
\centering
\includegraphics[width=.9\textwidth]{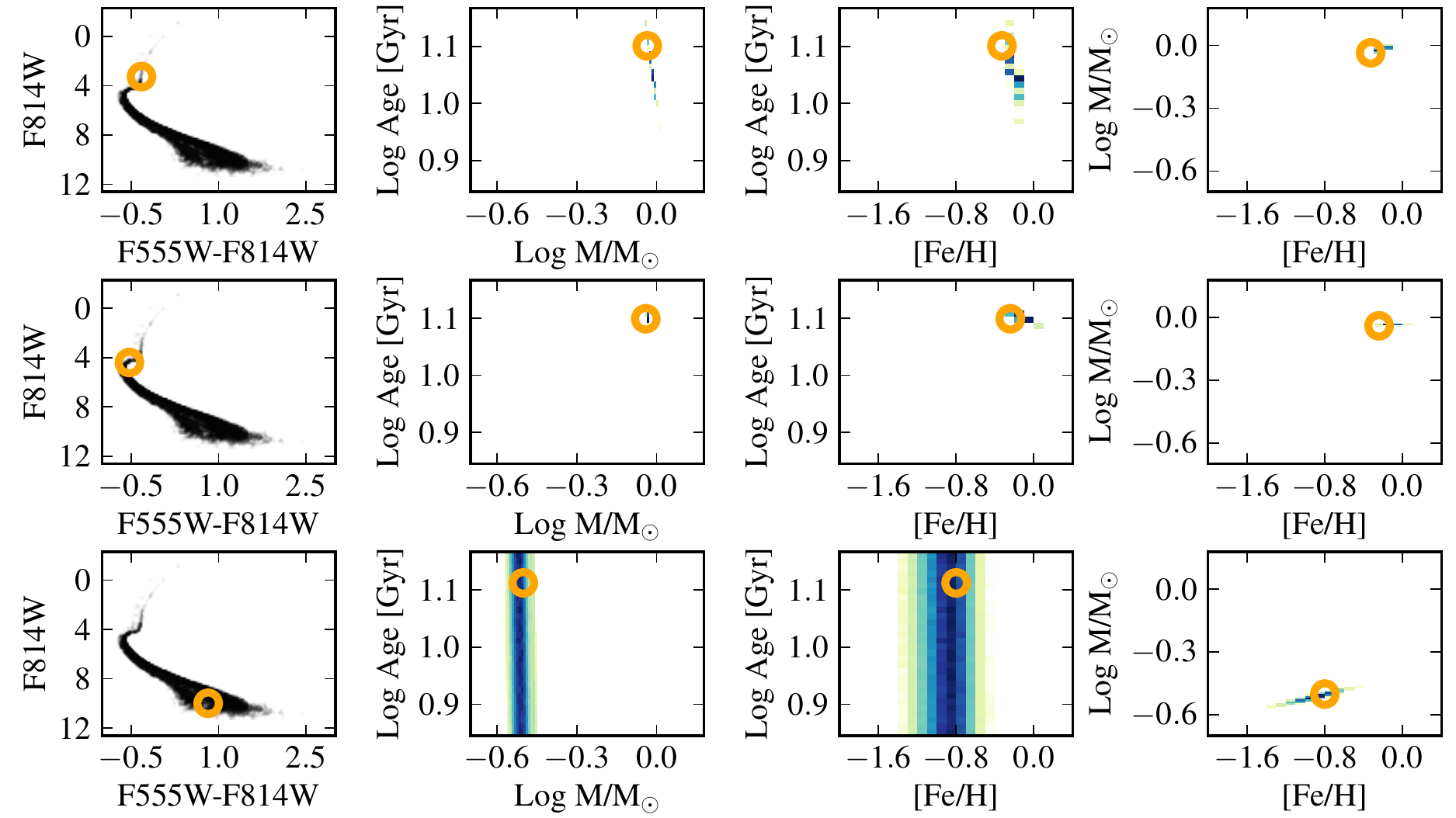}
\caption{Examples of likelihoods for stars in different evolutionary phases: RGB ({\it top row}), turnoff ({\it middle row}), and main sequence ({\it bottom row}). The left panels show the stellar location in the CMD. The other 3 panels show the likelihoods as functions of different pairs of variables, marginalized over the third. The orange circles indicate the observed magnitudes ({\it left column}) and the true mass, age, and metallicty values ({\it columns 2 -- 4}).
}
\label{fig:lkl2d}
\end{figure*}

Figure \ref{fig:lkl2d} shows how the stellar parameters are constrained differently in different stellar phases. The top row shows an RGB star with a mildly constrained age and very well constrained metallicity. The middle row shows a turnoff star for which the likelihood is very narrow in all directions. Finally, in the bottom row we show a main-sequence star, for which mass and metallicity are constrained moderately well, but for which age is almost completely unconstrained, at least within the range of ages adopted here. 

\section{Solving for the intensity function}
\label{sec:MCMC}

As detailed in Sect.~\ref{sec:summLambda}, the problem of solving for $\lambda$ can be regarded as solving for the IMF slope ($\gamma$) and the number of stars formed in each age-metallicity bin, $\{n_{i_{a,z}}\}$. Equation~(\ref{eq:finPPP}) describes the posterior pdf from which we wish to sample in order to obtain a solution. 
The general shape of star-formation histories in the age-metallicity plane is expected \emph{a-priori} to be sparse. This means that only a limited number of cells have occupancy $\{n_{i_{a,z}}\}$ greater than 0. In actual stellar fields, for example, the existence of an age-metallicity relation will greatly reduce the number of active model cells.
There are existing techniques, developed for image reconstruction, that allow one to deal with situations in which only a few pixels of the image contain the desired information while the rest of the pixels constitute a noisy background. 
These techniques allow one to suppress the background and sharpen the signal in the pixels where it is present. To improve our solution, we adopt an approach similar to that described in \cite{lingenfelter2009sparsity} and designed for Poisson-distributed data. 
This approach is equivalent to imposing a Lomax, or Pareto Type II, distribution as the prior (see Eq.~\ref{eq:finPPP}) on the pixel intensities ${n_{i_{a,z}}}$. The product of all of the cells' priors is proportional to
\begin{equation}
\label{eq:sparsity}
\prod_{i_{a,z}} \left(\frac{n_{i_{a,z}}}{\delta} + 1\right)^{-\beta}.
\end{equation}
In the language of regularization, $\beta$ sets the \emph{threshold}, or minimum value that is considered to be an actual, rather than spurious, signal, and $\delta$ sets the threshold's \emph{sharpness}, or strength with which values below the threshold are drawn towards zero.
After experimenting, we adopted $\delta = 2$ and $\beta = 0.4$, meaning a soft threshold, and meaning that we regularize (put to 0) only $(a,z)$ model cells with very low occupancy. 
At this value of $\beta$, the Lomax distribution is improper, i.e. integrates to infinity. Because the likelihood function of a PPP decays exponentially at high values of $\lambda$, the posterior pdf is still proper.

For the IMF slope $\gamma$, we assume a uniform prior bounded between $-3$ and $+3$. This range includes all of the commonly assumed high-mass IMF slopes.

Given the high dimensionality of this problem ($D = 776$ with the grid used here), sampling the whole space of $(\gamma,\,\{n_{i_{a,z}}\})$ can be inefficient, implying slow convergence when using traditional Markov Chain Monte Carlo sampling methods.
In order to accelerate convergence, we first estimate the maximum-a-posteriori (MAP) solution for $(\lambda, \{n_{i_{a,z}}\})$ using an Expectation-Maximization algorithm (see Appendix ~\ref{sec:emalg}).
We then start from the MAP estimate and use Metropolis-Hastings (M-H) Markov Chain Monte Carlo (MCMC) to generate samples from the posterior pdf.
M-H MCMC performs adequately and has the additional merit of being simple and intuitive.

\section{Simulated catalogs}
\label{sec:simcat}

\begin{deluxetable}{lclccccc}
\tabletypesize{\small}
\tablecaption{Main characteristics of the simulated catalogs\label{tab:catalogs}}
\tablewidth{0pt}
\tablehead{
\colhead{Name} & 
\colhead{$\gamma$ \tablenotemark{a}} & 
\colhead{$\Phi(a,z)$ \tablenotemark{b}} & 
\colhead{$N_{true}$} & 
\colhead{$N_{born}$} &
\colhead{$N_{obs}$} & 
\colhead{$f_{\mathrm{err}}$ \tablenotemark{c}} &
\colhead{$\sigma_{\mathrm{AMR}}$ \tablenotemark{d}}
}
\startdata
{\sc Blg}       & -2 & Bulge-like SFH and MDF           & 25000  & 25256 & 9921  & 1   & 0.05\\
{\sc Blg:No390} & -2 & Same as {\sc Blg}, no F390W data & 25000  & 24910 & 18480 & 1   & 0.05\\
{\sc Blg:F2p5}  & -2 & Same as {\sc Blg}, larger error  & 25000  & 25123 & 6264  & 2.5 & 0.05\\
{\sc Blg:Lrg}   & -2 & Same as {\sc Blg}, more stars    & 100000 & 100138& 39246 & 1   & 0.05\\
{\sc Burst3}    & -2 & Three equal-intensity bursts         & 25000  & 25002 & 10256 & 1   & 0.\\
{\sc Exp}       & -2 & Exponential decay            & 25000  & 25237 & 12298 & 1   & 0.\\ 
{\sc Const}     & -2 & Constant star formation          & 25000  & 25404 & 11624 & 1   & 0.\\
\enddata
\tablecomments{The {\sc BLG} catalogs have SFH and MDF that emulate observations of the Galactic Bulge. The {\sc Burst3} catalog has burst at $(a,z) = (12.5,-1.0); (8.5,-0.3); (8.5, -1.0)$. In both the {\sc Exp} and {\sc Const} catalog there is a linear age-metallicity relation with $a \in [10,12]$ Gyr and $z\in [-1.5,-0.5]$ (see also Figs.~\ref{fig:extcat5} and \ref{fig:extcat6}). The exponential decay is such that within the overall time interval, the star formation decays by 3 e-folds.}
\tablenotetext{a}{IMF slope}
\tablenotetext{b}{Combination of star-formation history and age-metallicity relation}
\tablenotetext{c}{Multiplication factor for the error curves with respect to the standard curve of Fig.~\ref{fig:errinc}}
\tablenotetext{d}{Dispersion about the mean age-metallicty relation}
\end{deluxetable}

To test our method, we simulate catalogs with a range of star-formation histories and age-metallicity relations.
We explore the impact of increased noise levels in the data and the impact of different magnitude cuts applied to the catalogs.
We also show how results differ when the metallicity-sensitive band $F390W$ is removed from the analysis.

For each catalog, we define a shape for $\lambda(m,a,z) \propto \mathcal{I}(m)\,\Phi(a,z)$.
We also specify the true number of expected stars ($N_{\mathrm{true}}$), equivalent to specifying the integral of $\lambda$. 
The actual number of stars formed ($N_{\mathrm{born}}$) is extracted from a Poisson distribution with expected value equal to $N_{\mathrm{true}}$.
Given the IMF single power-law parametrization, we need only specify the slope parameter; we set $\gamma = -2$ for all the simulated catalogs.
The function $\Phi(a,z)$ represents the age-metallicity relation for all the stars that are formed. We generate catalogs with isolated narrow bursts of star formation, as well as more extended star-formation histories, with large changes of stellar metallicity with age. 
Table~\ref{tab:catalogs} summarizes the characteristics of each catalog.

\begin{figure*}[!t]
\centering
\includegraphics[width=.99\textwidth]{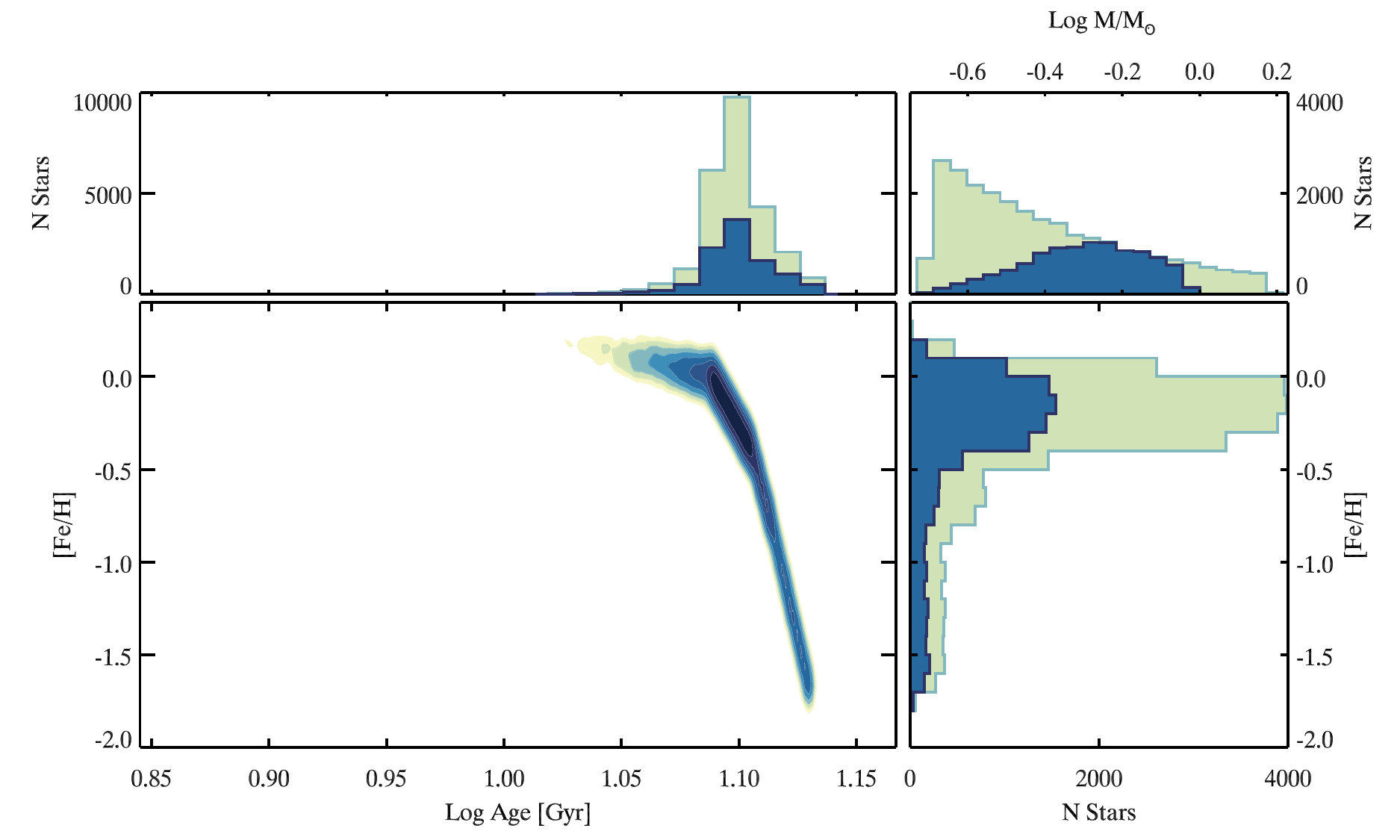}

\vspace{2pt}

\includegraphics[width=.96\textwidth]{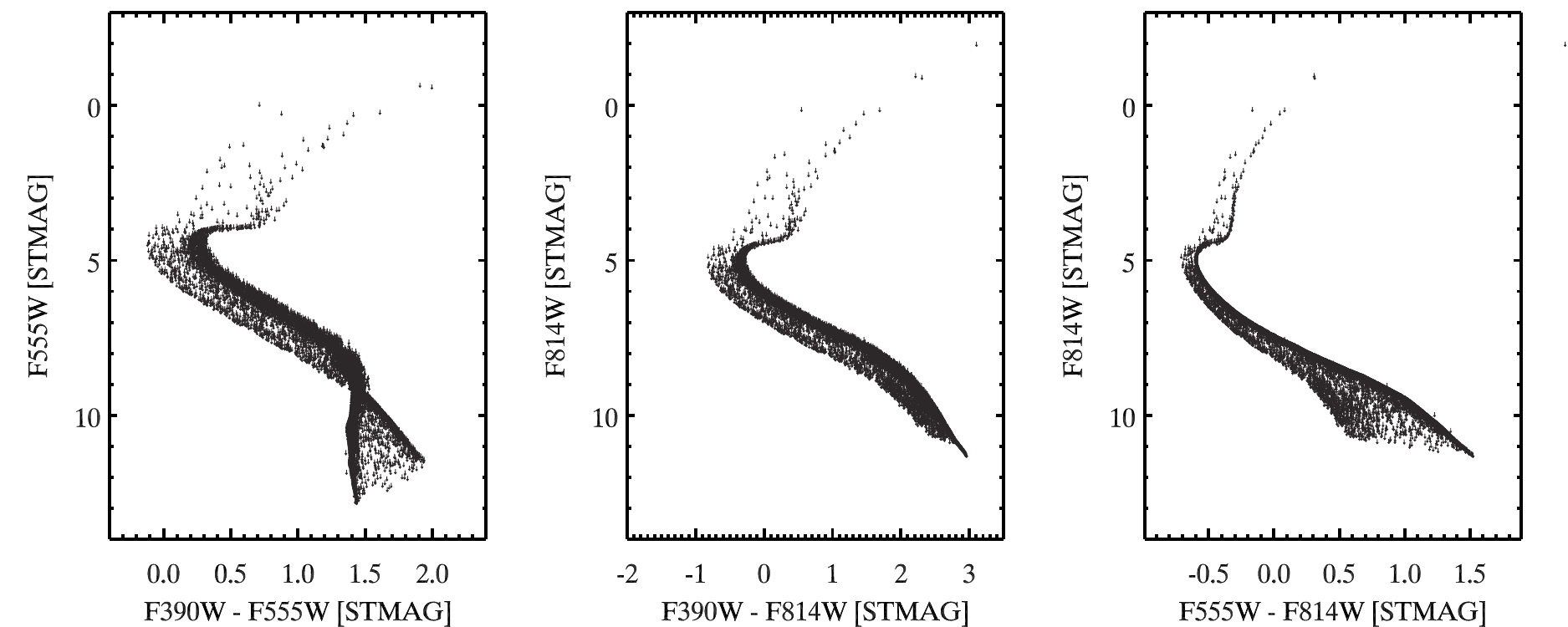}
\caption{Generated parameters (top) and simulated photometry (bottom) for the {\sc Blg} catalog. The contour plot shows the form of $\Phi(a,z)$. The three histograms show the mass, age and metallicity distribution for the whole simulated population (light blue) and the stars that make it to the simulated, observed catalog (dark blue). The latter are shown in three different CMD combinations in the bottom panels.\label{fig:extcat0} 
}
\end{figure*}

Each star's $(m,a,z)$ values are extracted from $\mathcal{I}(m)$ and $\Phi(a,z)$. Given the finite lifetime of stars, not all of the extracted values correspond to observable stars. Stars that are too massive for the extracted age and metallicity are still considered for the budget of formed stars but they obviously do not end up in the observable catalog.
If instead the extracted age does not exceed the maximum lifetime at given $m$ and $z$, stellar models are used to compute the intrinsic stellar magnitudes. 
We then use our noise model to assign each star a set of measurable magnitudes, and include or exclude the star from the final catalog based on the incompleteness curves in Fig.~\ref{fig:errinc}.
For some of the catalogs, we use a version of the error curves in Fig.~\ref{fig:errinc} where the whole curve is multiplied by a factor ($f_{\mathrm{err}}$) equal to 2.5, in order to explore the effects of reduced data quality. The completeness curves are also recomputed accordingly.
Figure~\ref{fig:extcat0} shows the extracted parameters for catalog {\sc Blg}. The histograms represent all of the formed stars (light blue) with the subset of observed stars superimposed (dark blue). The contour plot shows the form of $\Phi(a,z)$ chosen for this catalog. The three panels in the bottom row of Fig.~\ref{fig:extcat0} show the CMDs for the same catalog.
Analogous figures for the other catalogs are shown in Appendix~\ref{sec:catfig}.

\section{Results}
\label{sec:res}

\begin{figure*}[!t]
\centering
\includegraphics[height=7.5cm]{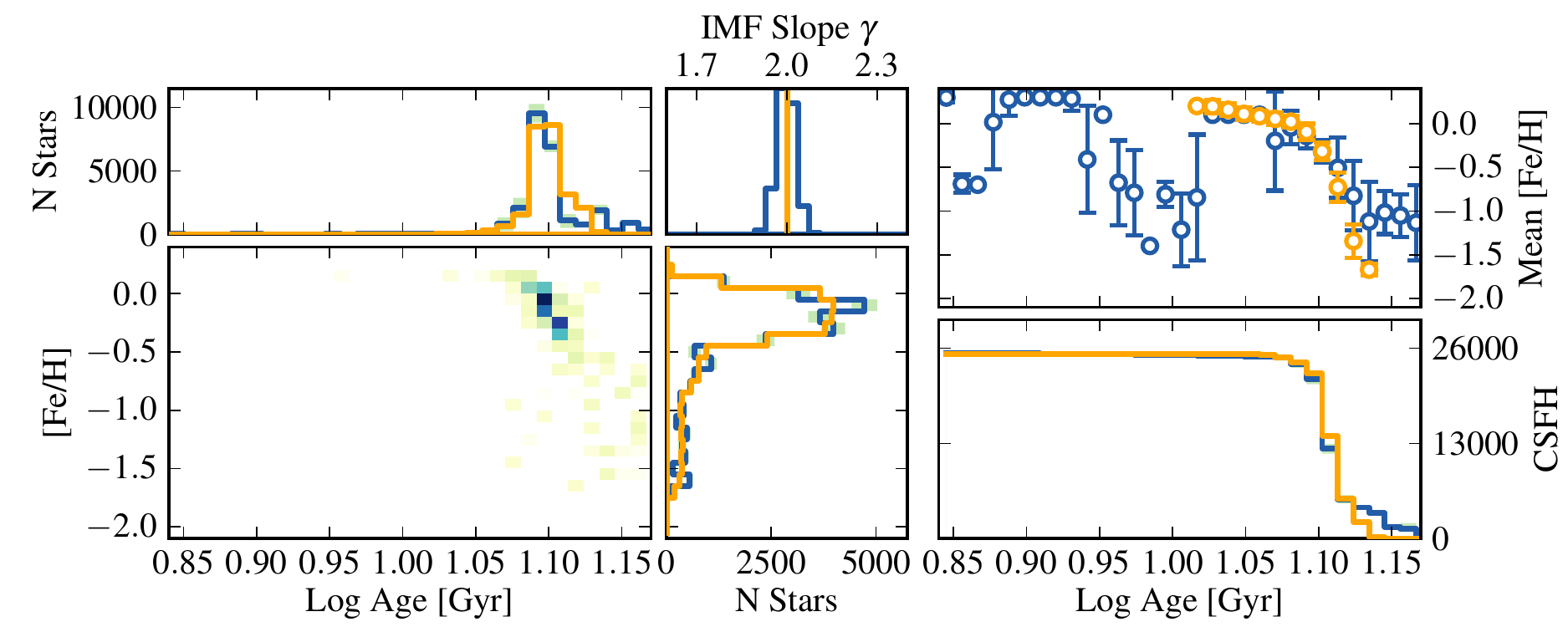}
\caption{Recovered properties for catalog {\sc Blg}. Left, bottom: the maximum-a-posteriori solution for $\{n_{i_{a,z}}\}$. Left, top: star-formation history. Center, bottom: metallicity distribution function. Center, top: distribution of the IMF slope values from the MCMC runs. Right, bottom: cumulative star-formation history. Right, top: mean age-metallicity relation and its dispersion. In all panels the simulated values are in orange, the recovered ones in shades of blue.\label{fig:rescat0_icut20} 
}
\end{figure*}

We use the outputs of the MCMC runs, Sect.~\ref{sec:MCMC}, to define our estimates of the parameters and their uncertainties. 
The results are summarized in Figs.~\ref{fig:rescat0_icut20}--\ref{fig:rescat_spec}.
In all the panels, different shades of blue represent the recovered solutions, while the original values are in orange. 

For the 2D star-formation history (left, bottom), we show the MAP solution as a way to represent the best combination of $\{n_{i_{a,z}}\}$ values.
Defining the uncertainty interval for the recovered solution in the age-metallicty plane is not straightforward. While we have samples for the $\{n_{i_{a,z}}\}$, it is the global solution for the whole (a,z) plane, and its uncertainty, that is of interest.
Adjacent cells in the (a,z) grid are strongly coupled by the fact that most of the individual stellar likelihoods are spread over large age-metallicity regions. This means that the pdf for the $\{n_{i_{a,z}}\}$ has a complex covariance structure, which is hard to represent and convey in a meaningful way.
Instead, we show the uncertainties of different collapsed 1-dimensional versions of $n_{i_{a,z}}$. 
We plot the marginal age (left, top) and marginal metallicity (center, bottom) distributions, as well as the cumulative star-formation history (CSFH) (right, bottom) and the mean age-metallicity relation (AMR) (right, top). 

For the first three of these collapsed distributions, we use the geometric medians as our best estimates.
The geometric median is a commonly used central tendency indicator; it is the multidimensional generalization of the common median.
In this case, we want to identify a median marginal SFH, or MDF, or CSFH, and each grid point where such distributions are computed constitutes a dimension.
We treat the  marginal pdf, for one MCMC iteration, as a point in an $m_{\rm{grid}}$-dimensional space, where $m_{\rm{grid}}$ is the number of grid points along the age or metallicity direction. When considering all the MCMC iterations, the geometric median is the point in the $m_{\rm{grid}}$-dimensional space that minimizes the sum of the (Euclidean) distances to all the other points.
To quantify the uncertainty on the marginal distributions, we consider each bin individually.
We take the difference between the bin value of the geometric median and the bin value of the distributions for each MCMC iteration. We then sort the differences and take the 16\% (84\%) quantiles; these quantiles are analogous to the typical $1-\sigma$ interval in case of a Gaussian distribution.

We define the mean AMR as the mean value of the metallicity of stars formed within one age grid cell. 
The AMR can have an intrinsic dispersion (as in the case of all the {\sc Blg}-type catalogs), meaning that at fixed age, a star can be formed with a range of metallicities, as in the case of non-instantaneous mixing of the ISM.
Given the possible intrinsic dispersion, in the summary plots we only show the comparison between the simulated AMR and the MAP solution. To avoid confusion, we do not show the additional dispersion introduced in the solution by the fact that each MCMC sample can have a slightly different mean AMR.
It is worth noting that in most cases the AMR solutions in the top-right panels depart significantly from the input AMR (orange points).
At a closer look, however, it is clear that the departure is limited to metallicities where the 2D solutions are generally negligible. For these metallicities the input AMR is not defined (no stars are formed in the input model), but the output mean AMR can still be computed. When looking simultaneously at the AMR and 2D solution, it is clear that the recovered AMR could be, in fact, truncated to metallicities where the 2D solution is significant. However we show the full derived AMR for the sake of completeness,

Finally, the IMF slope panel (center, top) shows the input value of (orange) and posterior pdf for the IMF slope $\gamma$.

Figure~\ref{fig:rescat0_icut20} shows the results for catalog {\sc Blg}. The distribution of the recovered IMF slope is centered on the true value, and the marginal SFH and MDF match the input.
Some deviation is observed at old ages for the SFH, CSFH, and AMR, 
most likely because few stars formed at these ages and even fewer were observed.
Moreover, the stars are formed at low metallicity; the isochrone colors
become increasingly degenerate at low metallicities, leading to individual stellar likelihoods that are less well-defined, and looser AMR constraints.

\subsection{Catalog cuts and photometric errors}

\begin{figure*}[!t]
\centering
\includegraphics[height=6.5cm]{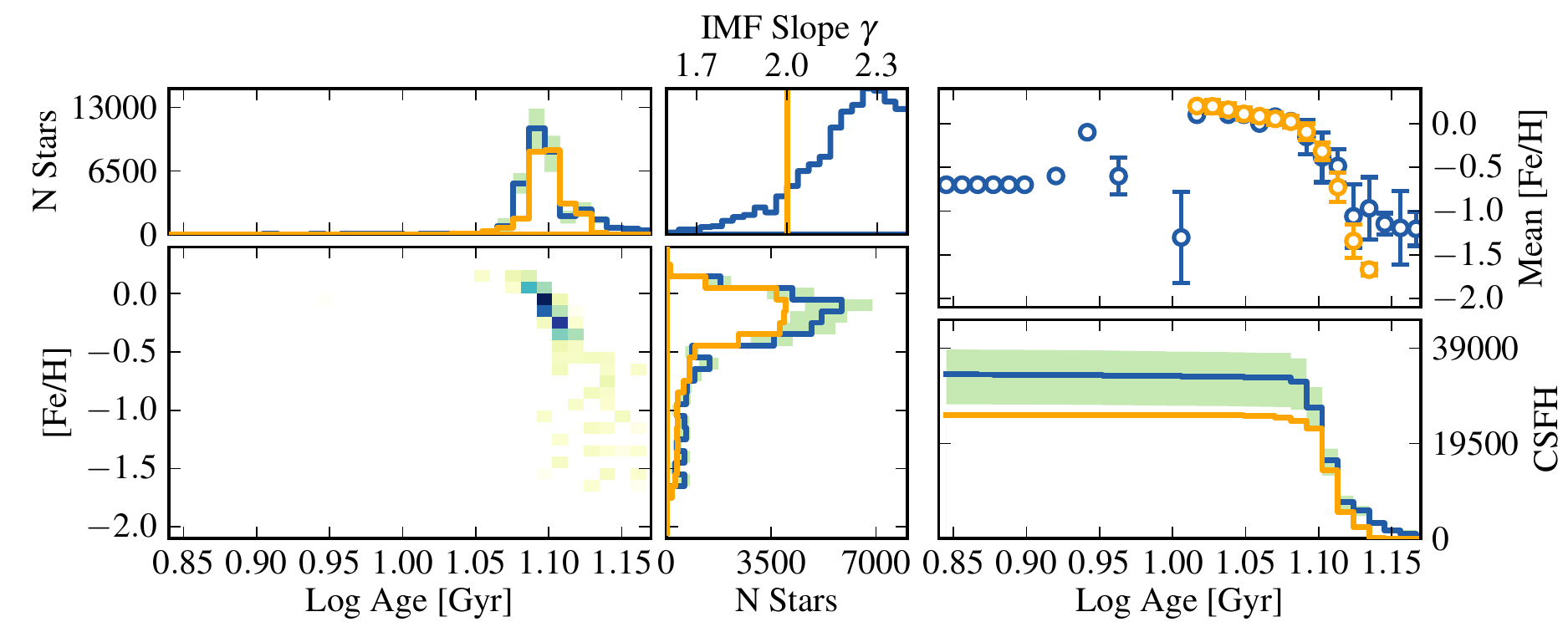}
\includegraphics[height=6.5cm]{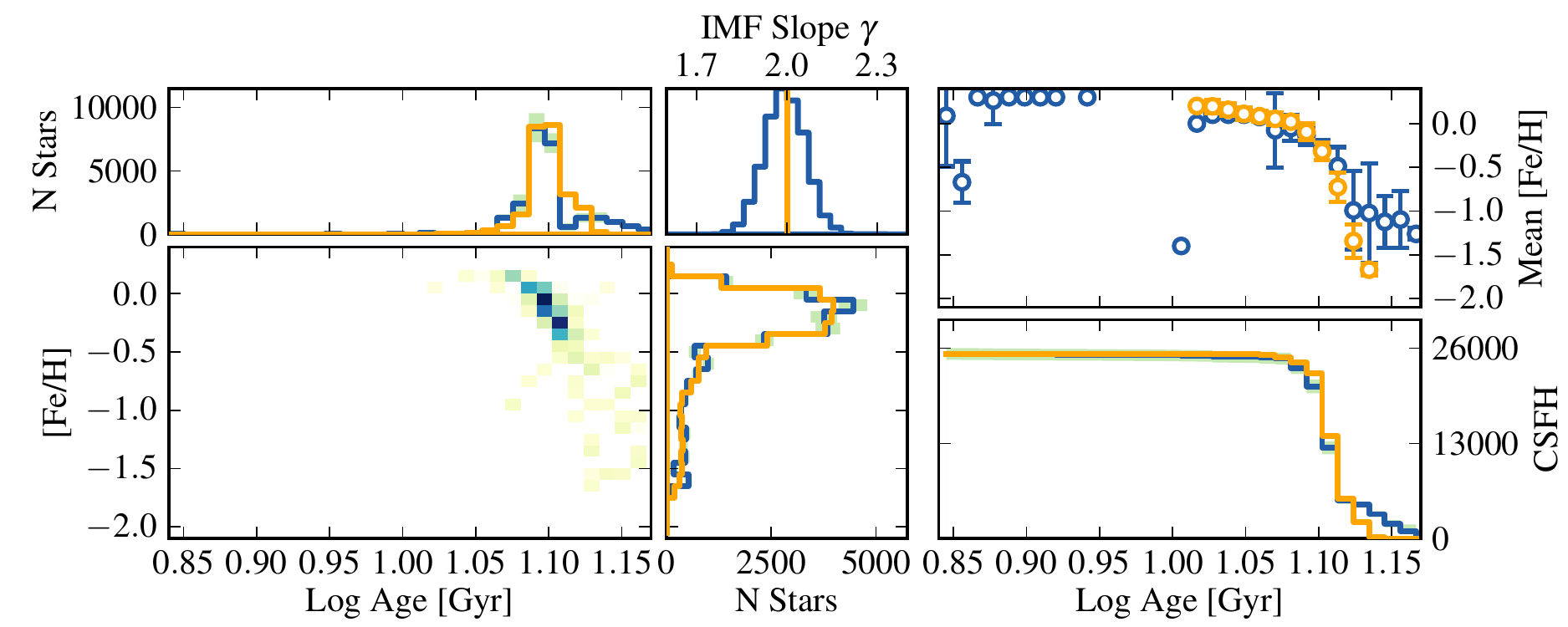}
\includegraphics[height=6.5cm]{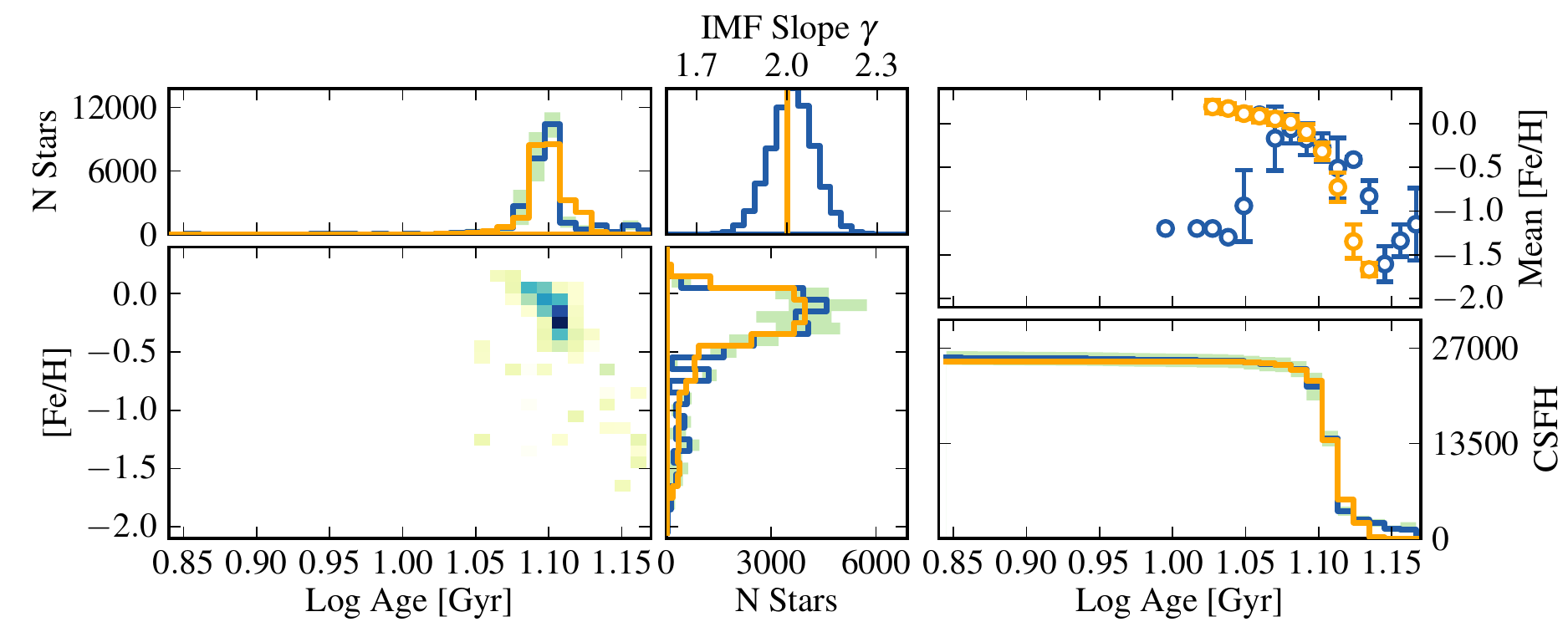}
\caption{\label{fig:rescat_0_i7_0_i9_2_i9} 
Recovered properties for catalog {\sc Blg}, with cut at $M_{814} = 7$ mag (top), $M_{814} = 9$ (center), and catalog {\sc Blg:F2p5}, cut at $M_{814} = 9$ mag (bottom). See Fig.~\ref{fig:rescat0_icut20}  for a description of the individual panels.}
\end{figure*}

In the case of Fig.~\ref{fig:rescat0_icut20} we have used all of the observed stars of catalog {\sc Blg} to reconstruct the intensity function.
However, as anticipated in Sect.~\ref{sec:compcomp}, there might be cases in which it is desirable to adopt conservative cuts, avoiding stars at the faint limit.
As illustrated in Sect.~\ref{sec:compcomp}, these hard cuts can be accounted for in the incompleteness function since they are simply another aspect of the detection process.
We show in Fig.~\ref{fig:rescat_0_i7_0_i9_2_i9} the effect of cuts at $M_{814} = 7$ (top) and 9 (center) mag, respectively. The $M_{814} = 9$ mag cut corresponds approximately to a cut at the 50\% completeness limit (see Fig.~\ref{fig:comp9}).

Such cuts affect different aspects of the solution in different ways. 
Generally speaking, the shape of the SFH, MDF, and AMR are not significantly changed. This is to be expected, since both of the applied cuts leave the turnoff and RGB intact. The turnoff carries most of the age information, while the RGB is a good metallicity indicator.
The cuts have some effect on the details of the SFH and MDF, especially in poorly-populated parts of the $(a,z)$ plane and where there are few, if any, turnoff or RGB stars. 
The largest difference is in the IMF estimates. The reduced mass range implies that the IMF slope cannot be recovered with the same accuracy and precision. This in turn affects the overall normalization of the SFH, as evident in the CSFH plot in the $M_{814} = 7$ mag cut case.
It is encouraging that the $M_{814} = 9$ mag cut case still looks very good, in all respects. Although the precision of the $\gamma$ recovery is lower than in the full catalog case, there is no obvious bias (in contrast to the $M_{814} = 7$ mag cut case). 

In an ideal situation, where one has perfect knowledge of the completeness and photometric errors at all magnitudes, it would be best to analyze the full catalog.  With real data, one must be careful using stars near the faint limit, because the effects of any biases in the characterization of the completeness or photometric uncertainties will be amplified.  Fortunately, a conservative cut to the catalog at the 50\% completeness limit, with proper inclusion of this cut in the selection function, can yield results that compare well to those from the full catalog.
For the rest of the examples, we will be mostly using catalogs with a $M_{814} = 9$ cut.

The bottom panel of Fig.~\ref{fig:rescat_0_i7_0_i9_2_i9} shows the results for catalog {\sc Blg:F2p5}, simulated with the same IMF, SFH, and MDF as catalog {\sc Blg}, but with photometric errors increased by a factor of 2.5; we limit the comparison to the case with $M_{814} = 9$ mag cut.
The overall recovery in this case is similar to the {\sc Blg} case. In particular, the IMF recovery looks very similar. This is because individual stellar masses are still very well recovered, at $M_{814} < 9$, because our mass grid spacing (2\%) is coarse enough that the {\sc Blg} and {\sc Blg:F2p5} cases are almost indistinguishable.
Similarly, the marginal (and cumulative) SFH and MDF are also well recovered. However, the details of the AMR start to get worse for the age bins with fewer stars; with the larger uncertainties, the information content of the few stars that were generated in those bins is no longer sufficiently constraining. 
This example demonstrates the need, in real cases, to understand the limitations of the available dataset.

\subsection{Impact of N$_{\mathrm{bands}}$ and N$_{\mathrm{obs}}$}

\begin{figure*}[!t]
\centering
\includegraphics[height=6.5cm]{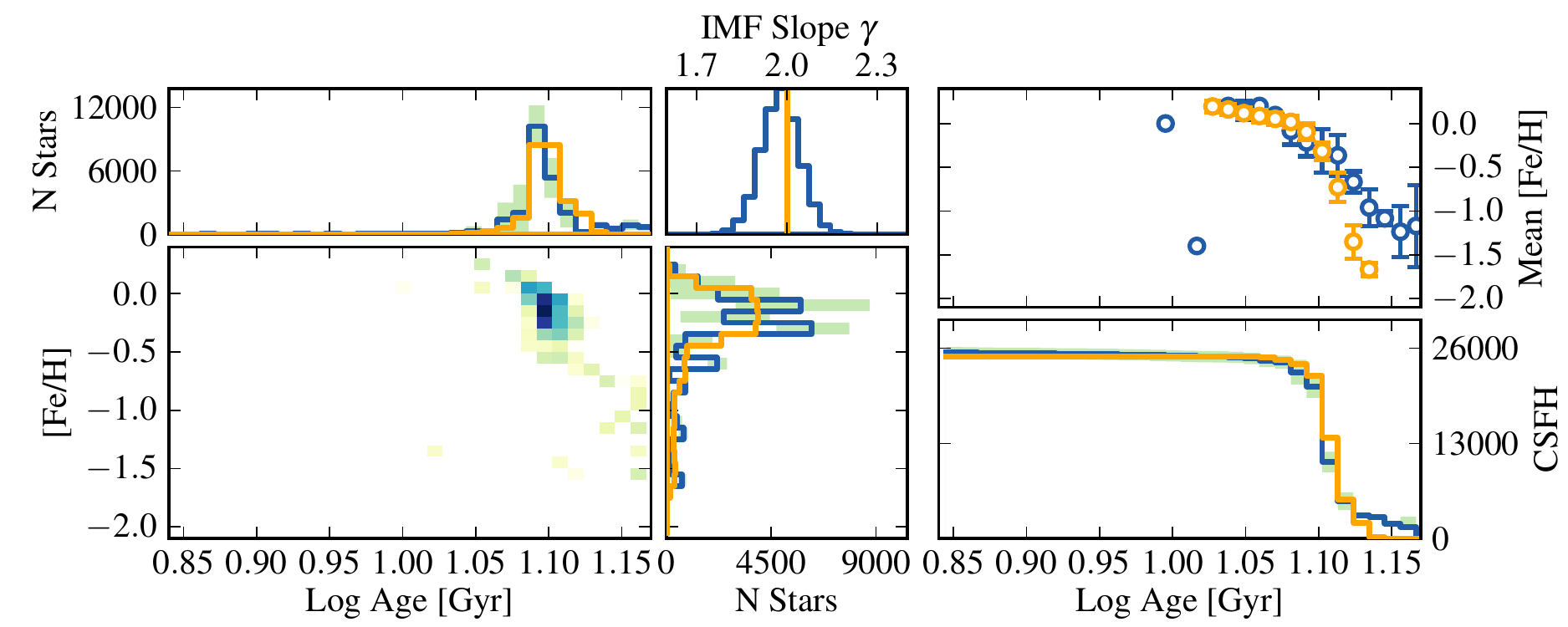}
\includegraphics[height=6.5cm]{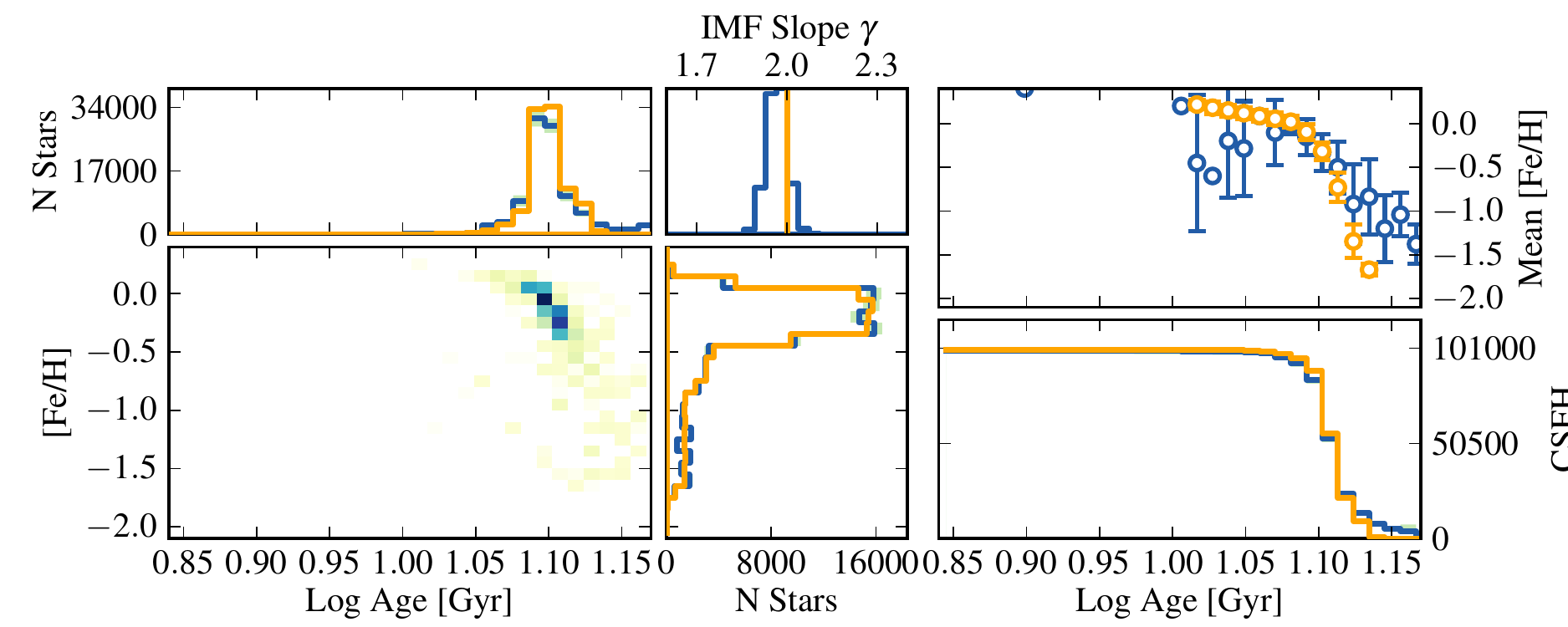}
\caption{\label{fig:rescat_1_i9_3_i9} 
Recovered properties for catalog {\sc Blg:No390} (top) and catalog {\sc Blg:Lrg} (bottom), both cut at $M_{814} = 9$ mag. See Fig.~\ref{fig:rescat0_icut20}  for a description of the individual panels.}
\end{figure*}

Catalog {\sc Blg:No390} has been simulated using the same IMF, SFH, and AMR as catalog {\sc Blg}, but  in this case we have changed the mock observing strategy to only consider 2 filters, omitting F390W. This choice enables deeper catalogs (see the number of stars in Table~\ref{tab:catalogs}) at the cost of having weaker metallicity constraints.
In our simulations, F390W is the shallowest band and has the worst signal-to-noise ratio.
This is a common choice when designing an observing strategy: more filters and a shallower catalog vs. deeper observations and less chromatic information. 
The top panel of Figure~\ref{fig:rescat_1_i9_3_i9} demonstrates that 
the deep observations guarantee a good result for the IMF slope, and hence the normalization of the SFH. The mean AMR, SFR, and MDF match the input to within the uncertainties. However, the MDF itself is highly uncertain in this observing configuration, as expected.

The bottom panel of the same figure shows the results for catalog {\sc Blg:Lrg}, whose $\Phi(a,z)$ is equal to 4 times that of {\sc Blg}; this corresponds to observing the same population over an area 4 times larger. The recovery of all quantities is nearly optimal.

\subsection{Differentiating similar histories}

In some real cases it might be very important to be able to distinguish between apparently similar star formation scenarios.
Catalogs {\sc Exp} and {\sc Const} have been designed to test the ability of our method to differentiate such cases.
The catalogs have the same IMF, AMR, photometric errors, and selection criteria, but the star-formation rate is exponentially decaying in one case and constant in the second. The total number of stars formed and the duration of star formation are also the same.
The scenarios' CMDs, which are shown in Figs.~\ref{fig:extcat5} and \ref{fig:extcat6}, appear quite similar.

\begin{figure*}[!t]
\centering
\includegraphics[height=6.5cm]{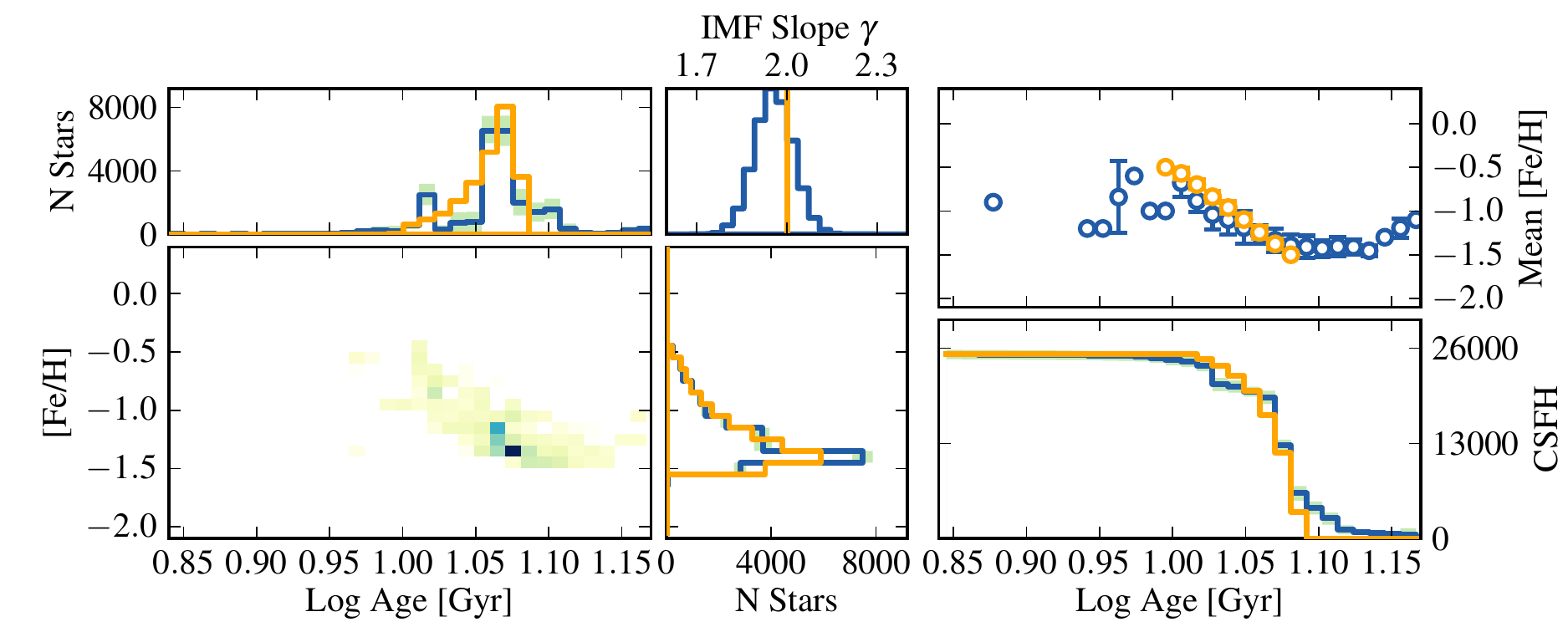}
\includegraphics[height=6.5cm]{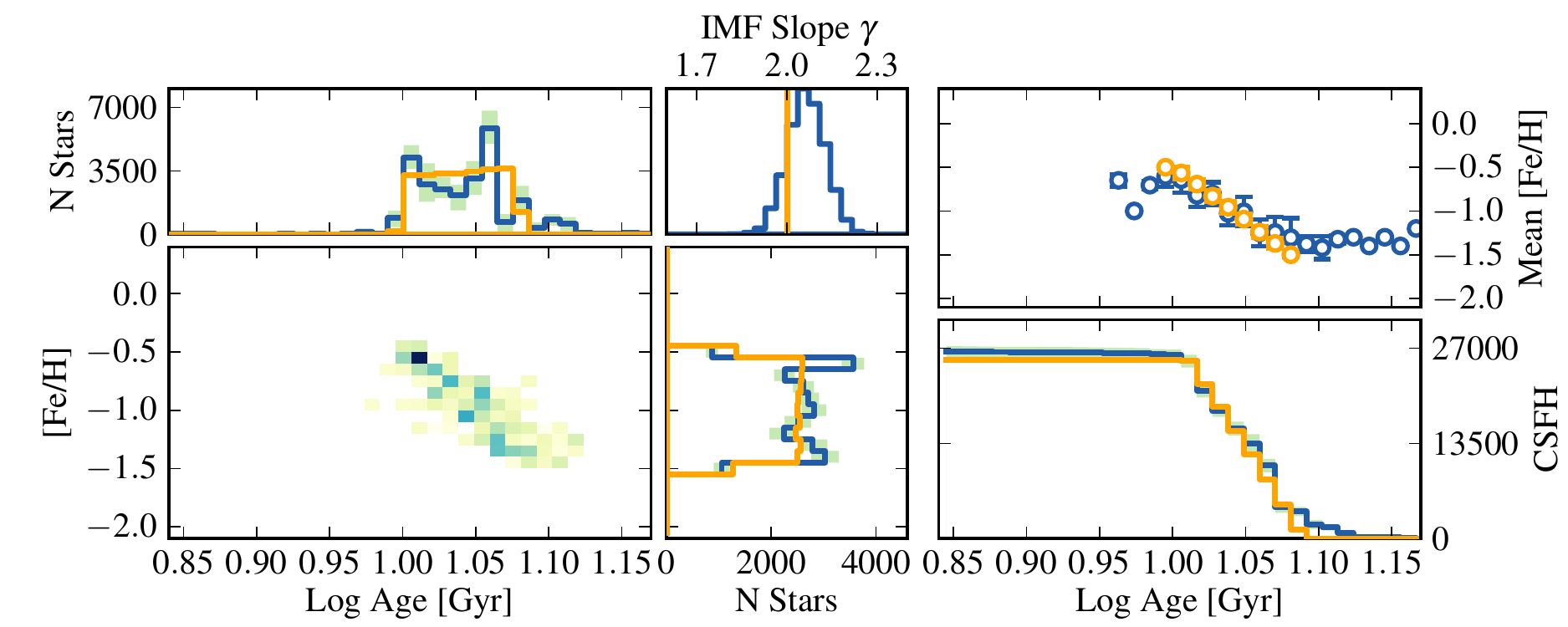}
\caption{\label{fig:rescat_5_i9_6_i9} 
Recovered properties for catalog {\sc Exp} (top) and catalog {\sc Const} (bottom), both cut at $M_{814} = 9$ mag. See Fig.~\ref{fig:rescat0_icut20}  for a description of the individual panels.}
\end{figure*}

The results for the recovery are shown in Fig.~\ref{fig:rescat_5_i9_6_i9}, for the cases in which both catalogs are cut at $M_{814} = 9$ mag. The recovery is again largely successful, with a clear distinction between the two scenarios. The differences are clearest in the CSFH plot.

\subsection{Nuisance parameters: the effect of a distance distribution}
\label{sec:nuipar}

As was explained at the end of Sect.~\ref{sec:summLambda}, our model can incorporate nuisance parameters (NPs) at the cost of increased computational time. 
We have tested the effects of introducing NPs by simulating catalogs with the same $\mathcal{I}(m)$ and $\Phi(a,z)$ as those in Tab.~\ref{tab:catalogs}, but with a distribution in distance modulus (DM). 
In particular, we will show results for the {\sc Burst3} catalog case, where three isolated peaks of star formation were simulated. This catalog provides a good visualization for the effects of a distance spread.
With respect to the original catalog, the one including a distance distribution has been spread using a Gaussian distribution in DM with expected value 0 and $\sigma_{\mathrm{DM}}= 0.25$ mag. This dispersion is similar to the spread in DM of stars along Galactic bulge sightlines.

As outlined at the end of Sect.~\ref{sec:summLambda}, when dealing with nuisance parameters both the incompleteness and likelihood function have to be computed as a function of the nuisance parameters too. A prior must then be specified in order to marginalize over them.

\begin{figure*}[!t]
\centering
\includegraphics[height=4.5cm]{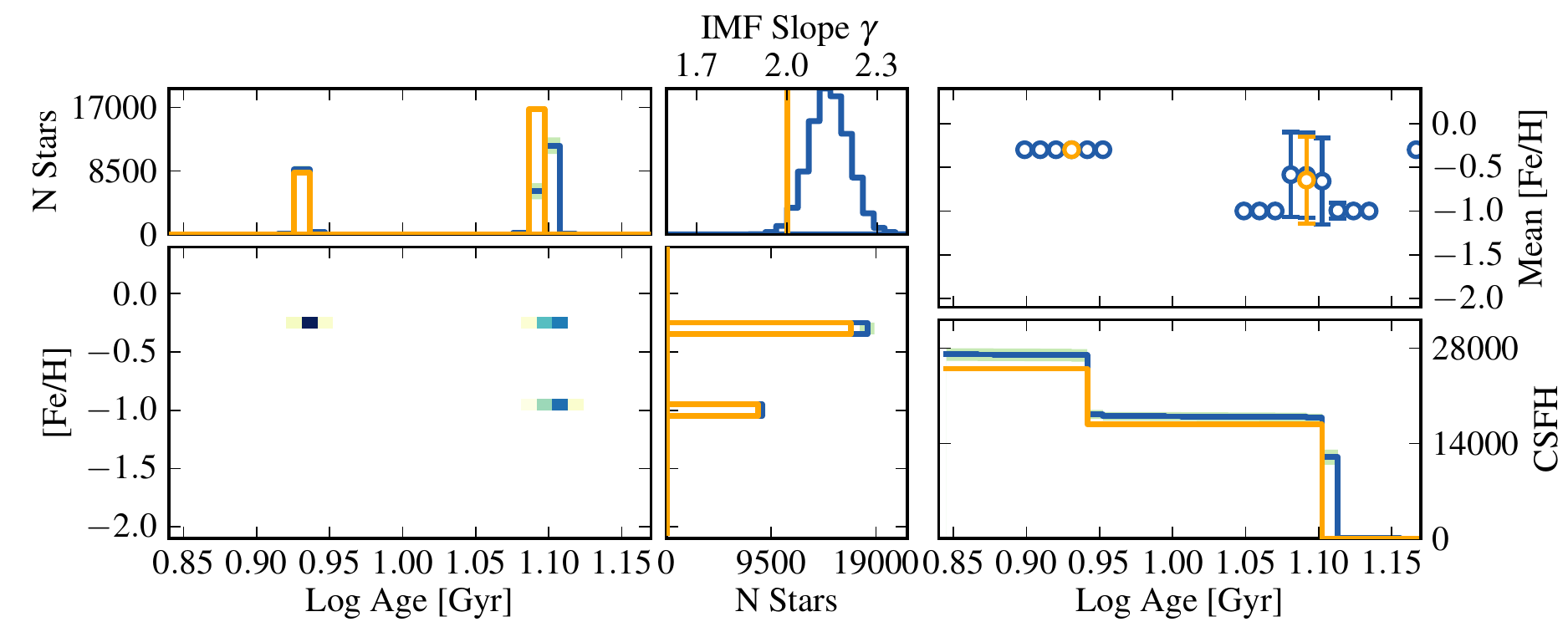}
\includegraphics[height=4.5cm]{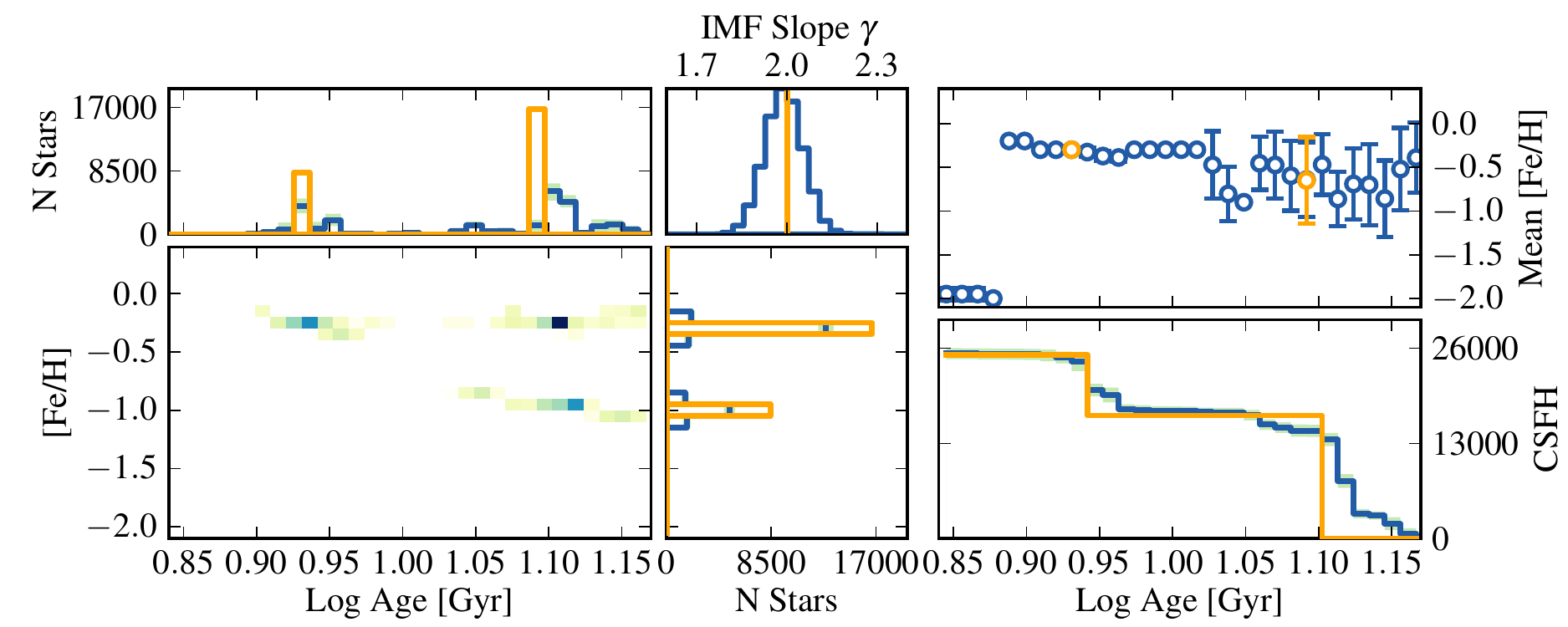}
\includegraphics[height=4.5cm]{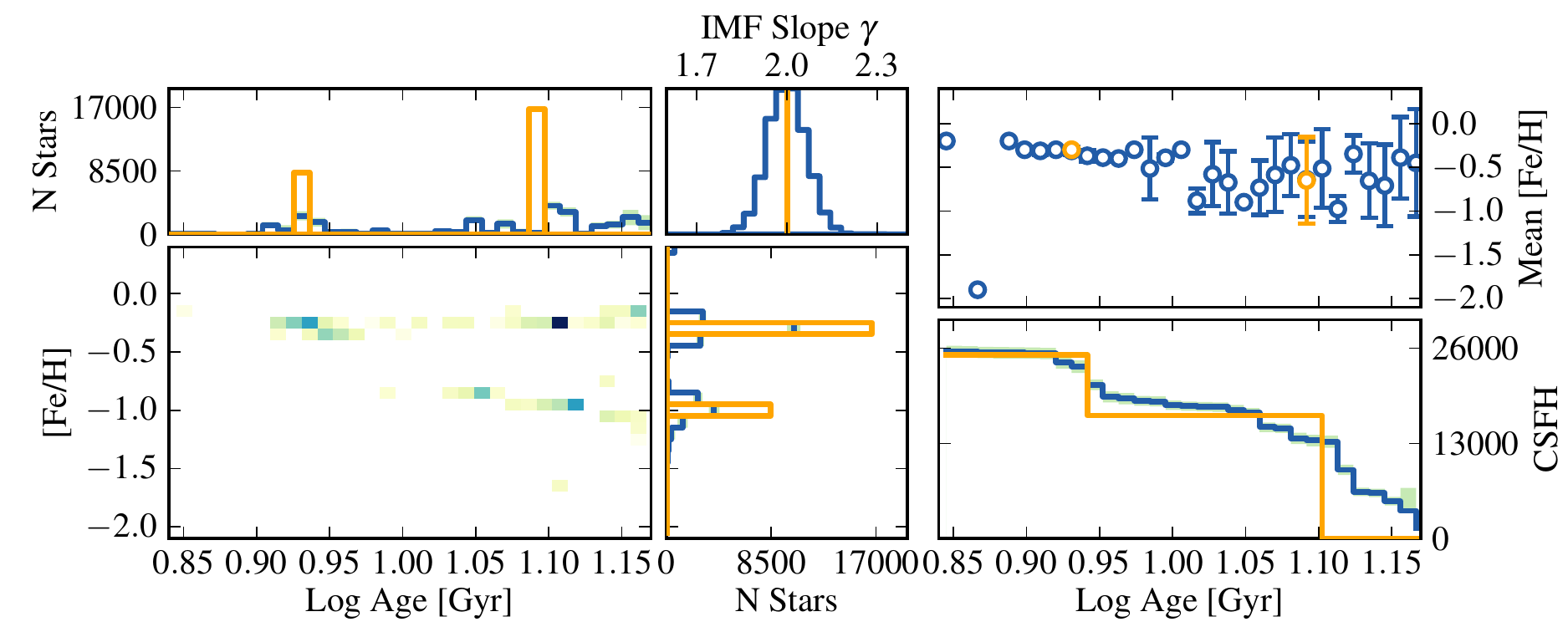}
\includegraphics[height=4.5cm]{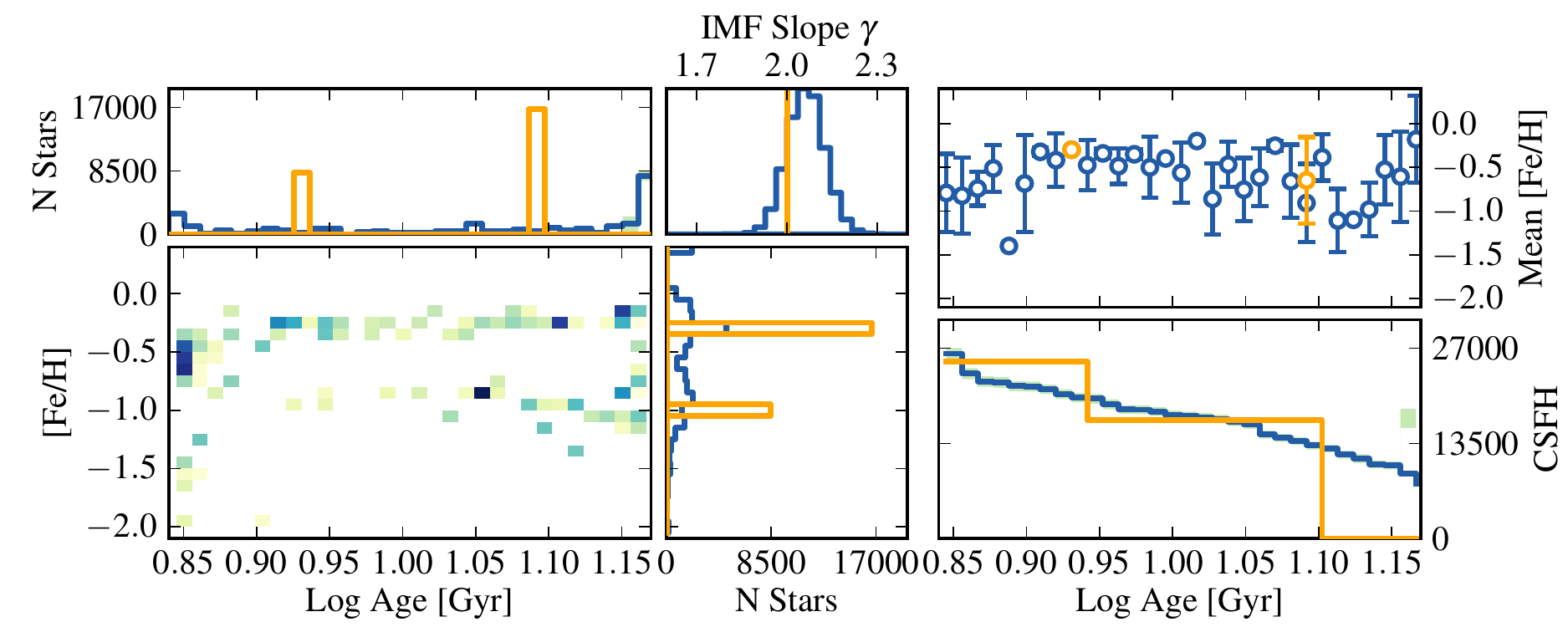}
\caption{\label{fig:rescat_4} 
Recovered properties for catalog {\sc Burst3}. Top: the case without distance spread. Second from the top: the correct distance distribution is used as distance prior in the recovery. Second from the bottom: a uniform distance prior between $-1$ and $+1 \sigma$ is used instead of the correct (Gaussian) one. Bottom: all the stars are considered to be $a-priori$ at the same distance, equal to the mean of the correct Gaussian prior. In all cases we adopted a catalog cut at $M_{814} = 9$ mag. See Fig.~\ref{fig:rescat0_icut20}  for a description of the individual panels.}
\end{figure*}

In Fig.~\ref{fig:rescat_4}, we show results where distinct priors are assumed for the distance
distribution (one correct and two incorrect).
Specifically, the results in the three panels are obtained under the following assumptions: a Gaussian prior with expected value 0  and $\sigma_{\mathrm{DM}}= 0.25$ mag (second from the top; the correct prior), a uniform prior with 0 mean and 0.25 mag width (second from the bottom; an incorrect prior that is still a reasonable approximation), and a single distance centered on 0 (bottom; an incorrect prior that is a
less reasonable approximation). 
For comparison, we also show the recovery in the case where all the stars are simulated at the same distance (top).

With a spread in distance, the degradation in the recovered IMF, AMR, SFH, and MDF is quite obvious.
Even in the case of the correct prior, the fact that the real distances of individual stars are known only with some probability makes both the individual ages and metallicities more uncertain. Essentially, all of the distributions have been convolved with the corresponding distance uncertainty. However, at least in the case with a correct distance prior (second from the top), the recovered values match the input to within uncertainties. In contrast,
incorrect assumptions on the distance prior further increase the discrepancies between the truth and the result.
In particular, assuming that all the stars are at the same distance makes all the answers for AMR, SFH, and MDF completely wrong while the IMF slope is less affected.

These examples constitute a serious warning against oversimplifying assumptions when it comes to priors.  In the case of the Galactic bulge, it is certainly true that the DM dispersion is
non-negligible, and so particular care should be taken regarding the assumed DM distribution.

\subsection{Applying spectroscopic constraints}

Often, in the study of resolved stellar populations, there are additional constraints that can be useful when solving for their IMF, SFH, and MDF. Spectroscopic constraints, usually from RGB stars, can improve the solution if they are appropriately handled.
As is the case for photometric incompleteness, it is very important to have a good knowledge of the selection function that is used to build the sample of targets for the spectroscopic observations.

A catalog of spectroscopic measurements and errors can be added to the catalog of photometric measurements to further constrain the intensity function ($\lambda$). 
Technically, the spectroscopic data constitute a new, related PPP, in the spectroscopic measurement space, with the same (up to normalization) underlying \emph{true} intensity as the one underlying the photometric measurements PPP.
The probability of the combined spectroscopic and photometric PPPs is the product of the probabilities of each one: 
\begin{multline*}
p(\{\eta_{\rm{spec}}\},  \{\eta_{\rm{phot}}\} \,| \,\lambda) = \\ p(\{\eta_{\rm{spec}}\}\,| \,\lambda, N_{\mathrm{spec}}) \times p(\{\eta_{\rm{phot}}\} \,|\, \lambda),
\end{multline*}
where $N_{\mathrm{spec}}$ is the (possibly different) normalization of the spectroscopic PPP.

The selection function for the spectroscopic sample can be regarded in the same way as the incompleteness function of photometric data and computed with the same methods described in Sect.~\ref{sec:compcomp}.
Once this function is evaluated, we can write an equivalent of Eq.~(\ref{eq:indlkl}) for the spectroscopic PPP. 

We simulate the process of building a spectroscopic sample and obtaining corresponding measurements by first generating a photometric catalog with the same true properties as the
{\sc Blg} catalog, but a larger total number of stars -- as noted above, the normalizations of the spectroscopic and photometric PPPs do not have to be same. We then select RGB stars with $2 < M_{814} < 3$ mag. This is analogous to selecting targets from a shallow wide-area survey centered on a similar position as the field where deep imaging is available for IMF, SFH, and MDF reconstruction. We then assign to each spectroscopic target an [Fe/H] error, extracted from a gamma distribution with shape parameter 50 and scale parameter 0.001, thus obtaining a mean error of 0.05 dex, and a standard deviation of the errors 0.007 dex. Finally, for each star we extract the measured [Fe/H] from a Gaussian centered on its true [Fe/H] value with $\sigma$ equal to the assigned error.  In the recovery, the individual likelihoods are independent of mass and age, as they only depend on [Fe/H]; the likelihoods are Gaussians centered on the observed metallicity with $\sigma$ given by the individual [Fe/H] measurement errors.  Here and in real datasets, the impact on computing resources is small, because the number of stars in the spectroscopic sample is much lower (generally only a few hundreds) than the photometric sample, and all the terms needed to compute the equivalent of Eq.~(\ref{eq:indlkl}) for the spectroscopic PPP are already calculated when solving for the photometric PPP.

\begin{figure*}[!t]
\centering
\includegraphics[height=4.5cm]{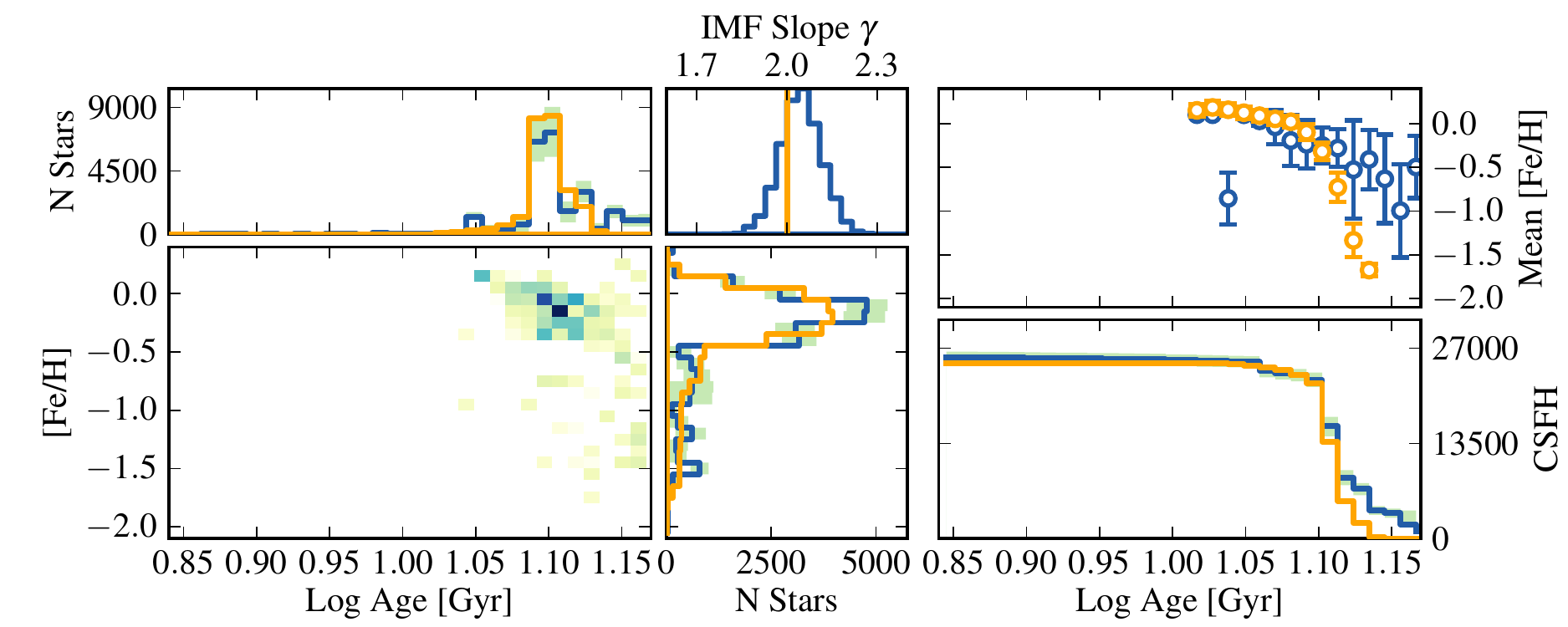}
\includegraphics[height=4.5cm]{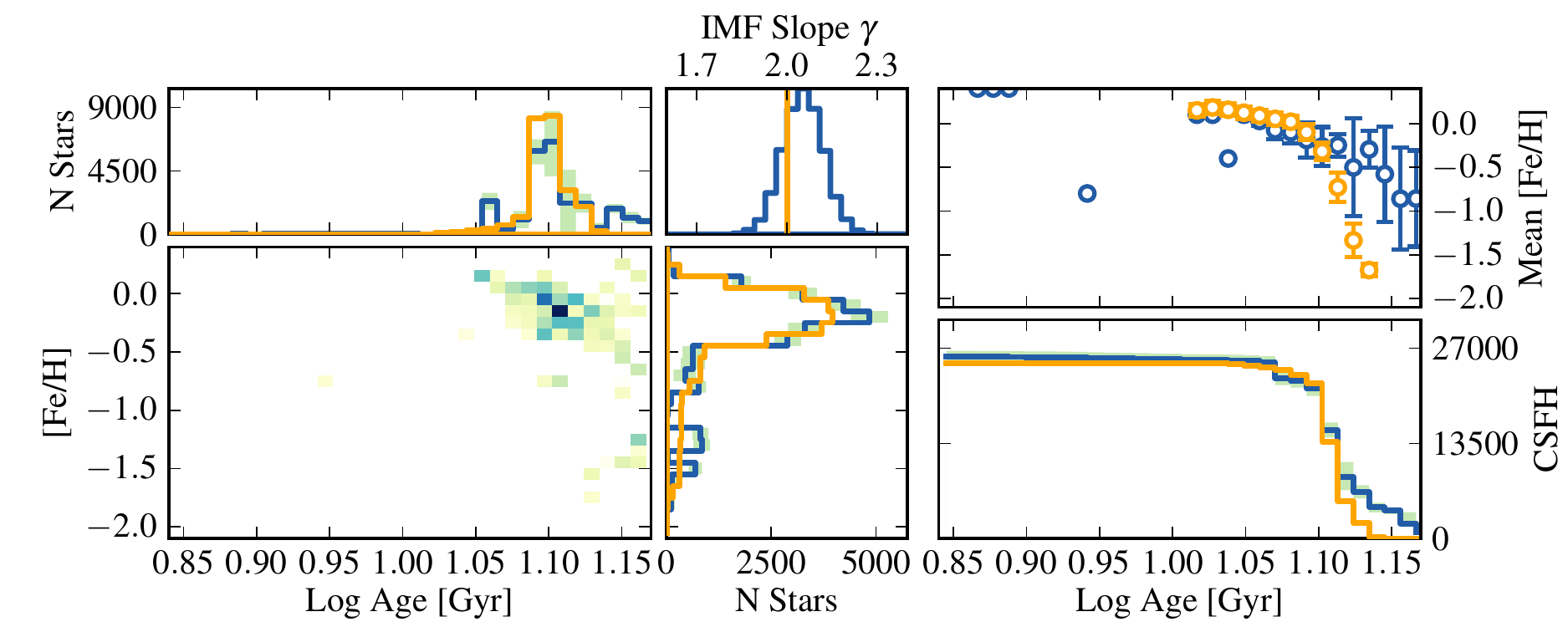}
\includegraphics[height=4.5cm]{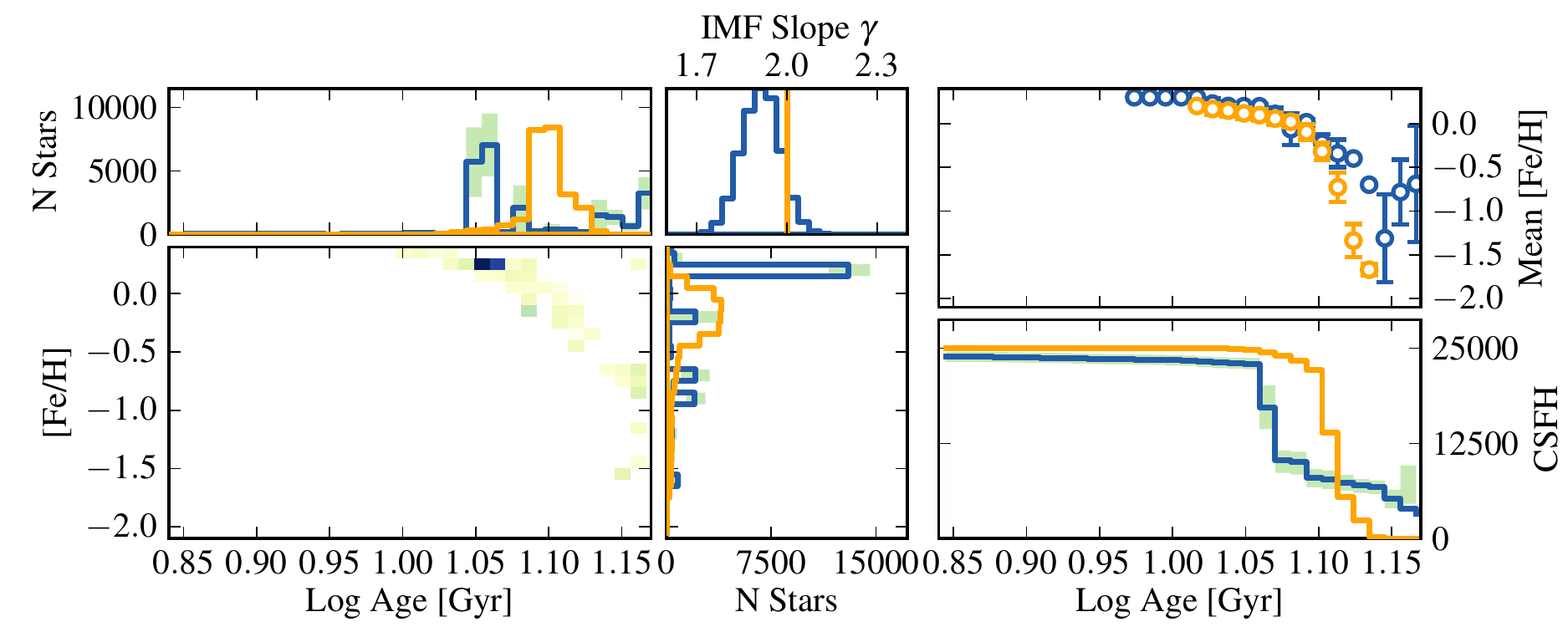}
\includegraphics[height=4.5cm]{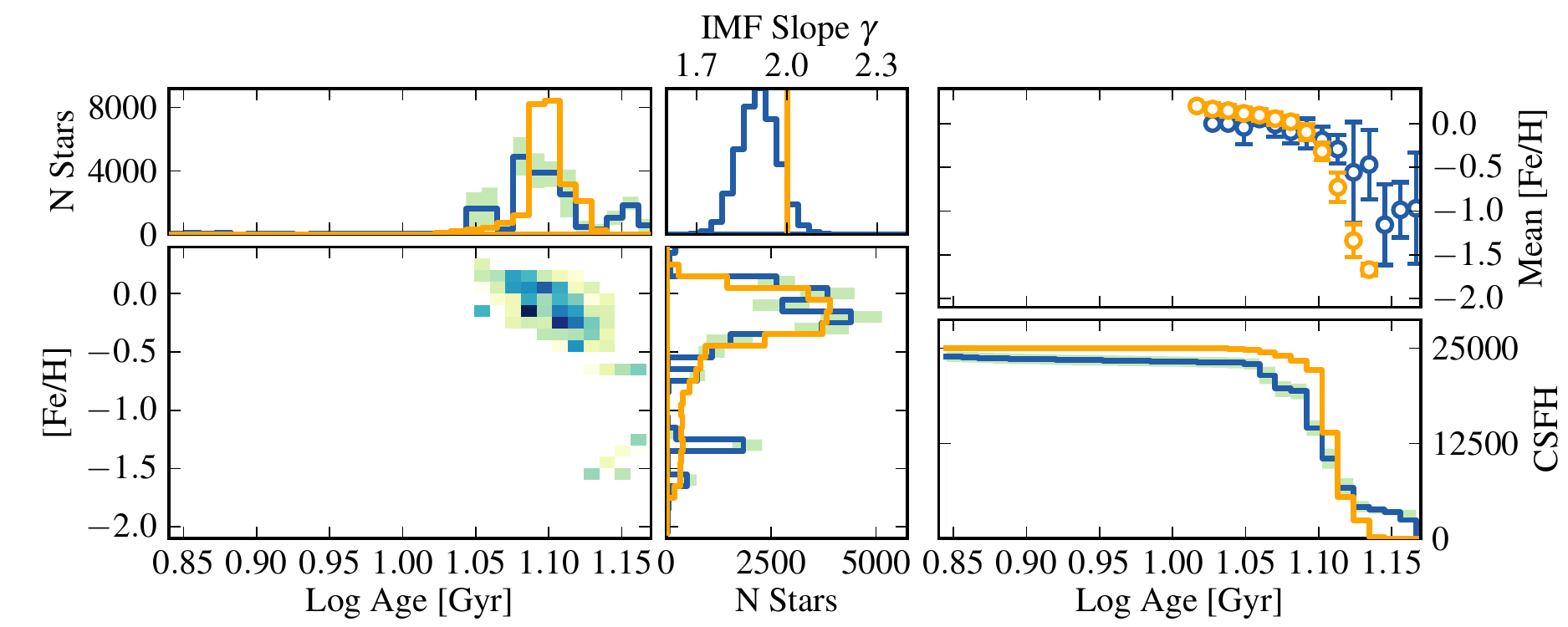}
\caption{\label{fig:rescat_spec} Results for catalogs {\sc Blg} (top two panels) and {\sc Blg:No390} (bottom two panels) with a Gaussian DM distribution. The first and third panels from the top show the results for the basic recovery. The second and fourth panels from the top show the results when a spectroscopic constraint on the MDF is applied.
 See Fig.~\ref{fig:rescat0_icut20}  for a description of the individual panels.}
\end{figure*}

To demonstrate the effects of a spectroscopic constraint, we show the cases of catalogs {\sc Blg} and {\sc Blg:No390}; the former is used as template, while the latter does not contain measurements for the metallicity-sensitive F390W filter. 
We put ourselves in the situation where the catalog is generated with a Gaussian DM distribution, with 0 mean and $\sigma = 0.25$ mag, and we use the correct prior to marginalize over distance. The results are shown in Fig.~\ref{fig:rescat_spec}.
In the first and third panels from the bottom, we show the results obtained without imposing a spectroscopic constraint, while in the second and fourth panels from the top the constraint is used. 
The results that include spectroscopy are more accurate and precise than the results that do not, though they are still not as good as the corresponding cases without a distance spread; this outcome is not surprising, given the examples of catalog {\sc Burst3} that explored in Sect.~\ref{sec:nuipar}. 
The improvement is more noticeable for catalog {\sc Blg:No390} (third and fourth panels from the top) than for catalog {\sc Blg} (first and second panels from the top), since the latter includes the metallicity-sensitive F390W band.

\section{Summary}
\label{sec:summary}
We have introduced a new approach to the study of resolved stellar populations via multi-band photometric observations. The outlined framework is based on PPP theory. 
We solve the problem using standard Markov Chain Monte Carlo techniques, combined with techniques developed for medical imaging reconstruction, such as sparsity regularization for Poisson Data.

The underlying idea driving this work was the need to simultaneously solve for the IMF slope, SFH, and MDF for nearby environments such as the Galactic bulge and the Milky Way satellites. We have developed a framework that allows easy inclusion of nuisance parameters, such as stellar distance, and demonstrated the importance of specifying informative priors for these nuisance parameters.
We have shown how to robustly incorporate measurement errors, incompleteness, and selection functions within the PPP framework. Our approach is particularly useful when multi-band data are available; in this case methods based on CMD gridding can be less straightforward to apply.
Another advantage of our approach is the ease with which we can incorporate certain types of additional observations, such as those coming from independent spectroscopic observations.

We have validated our method by simulating catalogs with different underlying stellar birth functions (number of stars formed per unit mass, age, and metallicity) and showing the how well we can recover the input values. We have tested the outcomes under different assumptions on the photometric errors, catalog size, selection function, available photometric bands, and accuracy of prior assumptions on the nuisance parameters.
These tests demonstrate that our technique recovers the input parameters without significant biases, limited only by the uncertainties in the data.

\bibliography{biblio_Gennaro_PPP}
\bibliographystyle{apj}

\appendix

\section{The Expectation-Maximization Algorithm}
\label{sec:emalg}

The Expectation-Maximization Algorithm (EMA) is a general iterative algorithm for finding the maximum-likelihood, or in our case maximum-a-posteriori (MAP), parameters of a probability distribution function when a direct solution is non-trivial.
There are many applications of the EMA; we refer the reader to \cite{MLalg}, \cite{EM} and \cite{Streit} for further details.
We summarize, without derivation, the EMA steps for PPPs on a discrete space (our grid of models).

Consider a set of $t_j, \, j= 1...M$ measurements (stellar magnitudes) and a piece-wise constant PPP, i.e. $\lambda(s) = \sum_{r=1}^K \lambda_r I_r(s)$, where $K$ is the number of grid cells, and $I_r$ is 1 across the r-th cell and 0 elsewhere. The likelihood of the data, given $\lambda$, is: 
\begin{equation}
\label{eq:likEM}
p(\{t\}\,|\,\lambda) = e^{-\sum_{r=1}^K \lambda_r |\mathcal{R}_r|} \prod_{j=1}^M \left(\sum_{r=1}^K \lambda_r\, f_r(t_j) \right),
\end{equation}
where
$$ 
\quad f_r(t_j) = \int p(t\,|\,s) \,I_r(s)\, \mathrm{d}s \quad \mbox{and} \quad |\mathcal{R}_r| = \int I_r(s)\mathrm{d}s.
$$

The idea of the EMA is to introduce a number of latent variables, usually referred to as missing data. In this case the missing data are the \emph{true} values of the stellar parameters, which we indicate with $u$. The joint probability of data and latent variables is given by:
\begin{equation}
\label{eq:joint}
p(\{t\}, \{u\}\, , \,\lambda) = e^{-\sum_{r=1}^K \lambda_r|\mathcal{R}_r|} \left( \prod_{j=1}^M \left(\lambda_{u_j} \, f_{u_j}(t_j) \right) \right) p(\lambda)
\end{equation}

Combining Eqs.~(\ref{eq:likEM}) and (\ref{eq:joint}) it is possible to derive the expression for the probability of the latent variables conditional on the data and $\lambda$:
\begin{equation}
\label{eq:cond}
p(\{u\} \,|\,\{t\}, \lambda) = \frac{p(\{u\},\{t\} \,|\, \lambda)}{p(\{t\}\,|\,\lambda)} = \prod_{j=1}^M \frac{\lambda_{u_j} \, f_{u_j}(t_j)}{\sum_{r=1}^K \lambda_r\, f_r(t_j) }
\end{equation}

The E-step consists of taking the logarithm of the joint probability, Eq.~(\ref{eq:joint}), and calculating its expectation value over the conditional, Eq.~(\ref{eq:cond}).
The M-step consists of maximizing the expression obtained in the E-step.
Without showing it, we note that the M-step requirements lead to an iterative scheme for computing $\lambda_r$ until a given convergence tolerance is reached, e.g. until the difference between the log-likelihood at step $n$ and $n+1$ is below a fixed threshold. This algorithm falls in the class of Shepp-Vardi algorithms \citep{shepp:vardi}.
Further modifications to the EMA are necessary when including the sparsity regularization penalty logarithmic function. We will not illustrate those modifications here and refer the reader to \cite{lingenfelter2009sparsity}, where the recursive updates for the EMA are derived for this particular case.

\section{Catalog Figures}
\label{sec:catfig}

\begin{figure*}[!t]
\centering
\includegraphics[width=.99\textwidth]{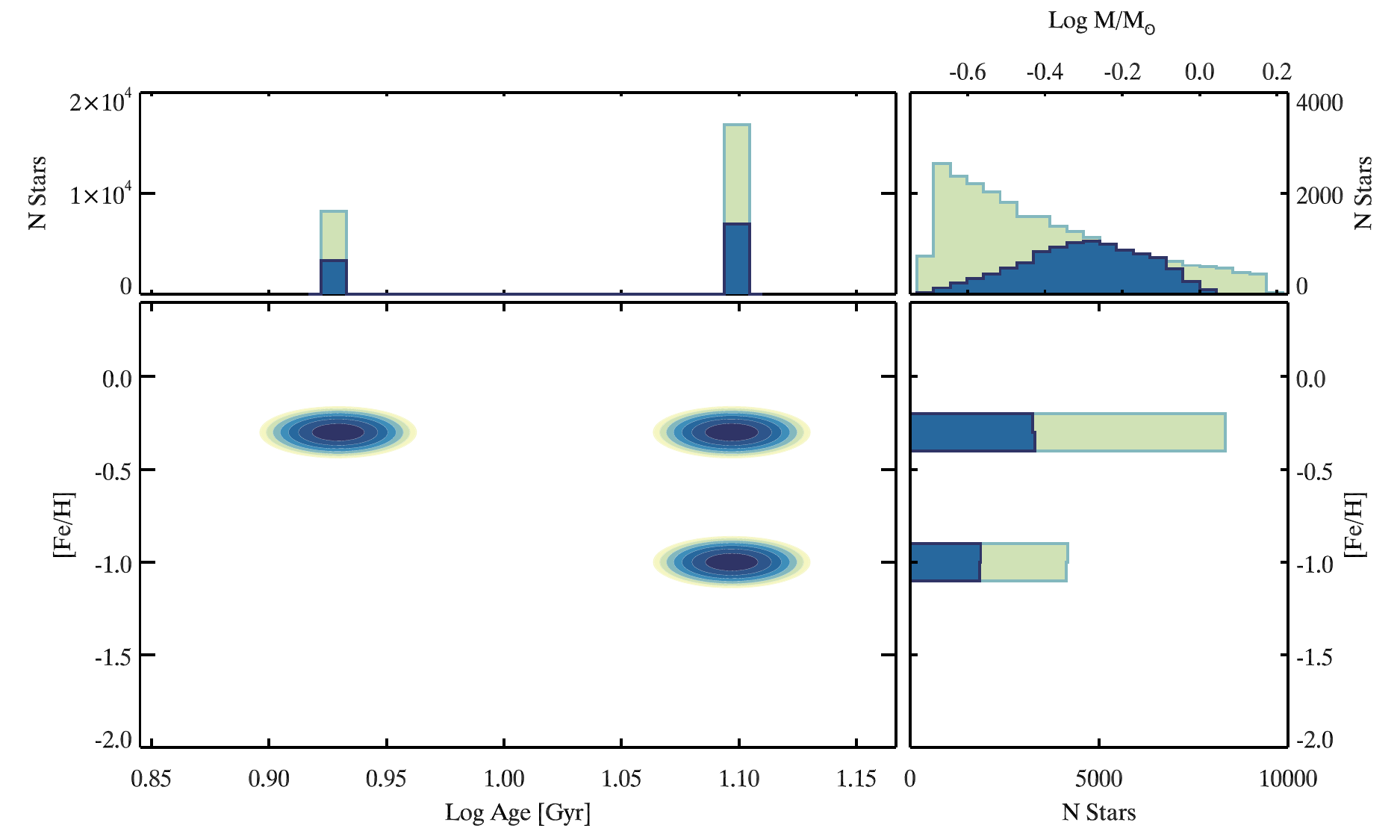}

\vspace{2pt}

\includegraphics[width=.96\textwidth]{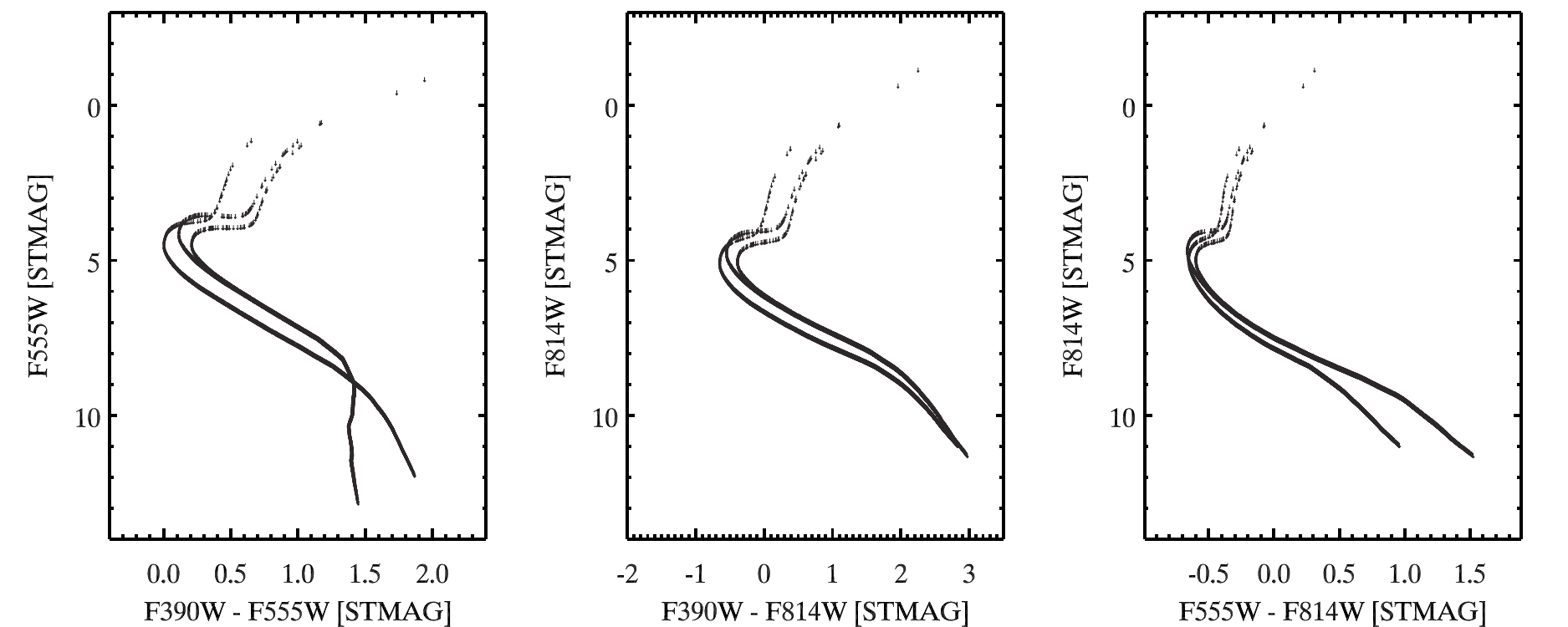}
\caption{Same as Fig.~\ref{fig:extcat0} but for the {\sc Burst3} catalog \label{fig:extcat4} 
}
\end{figure*}

\begin{figure*}[!t]
\centering
\includegraphics[width=.99\textwidth]{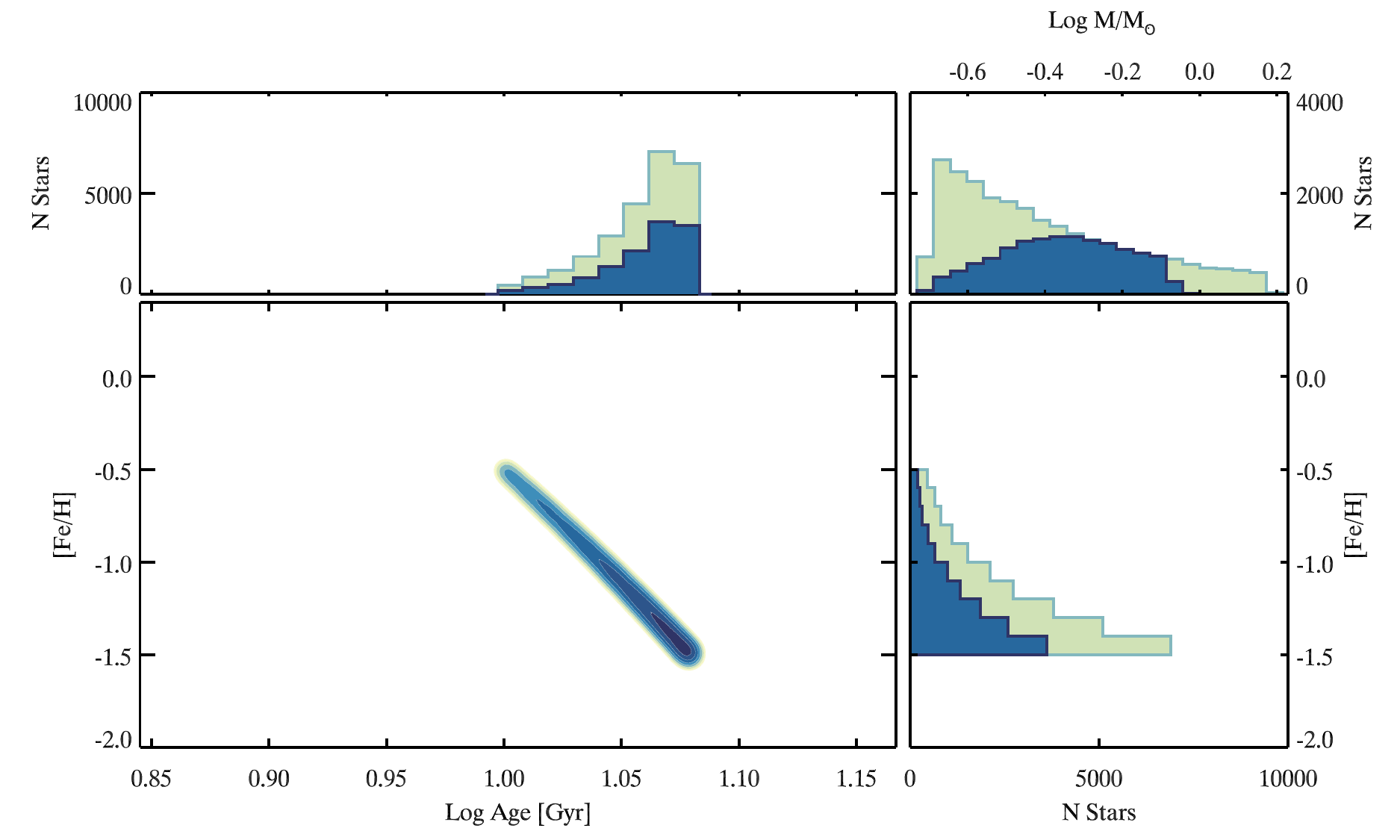}

\vspace{2pt}

\includegraphics[width=.96\textwidth]{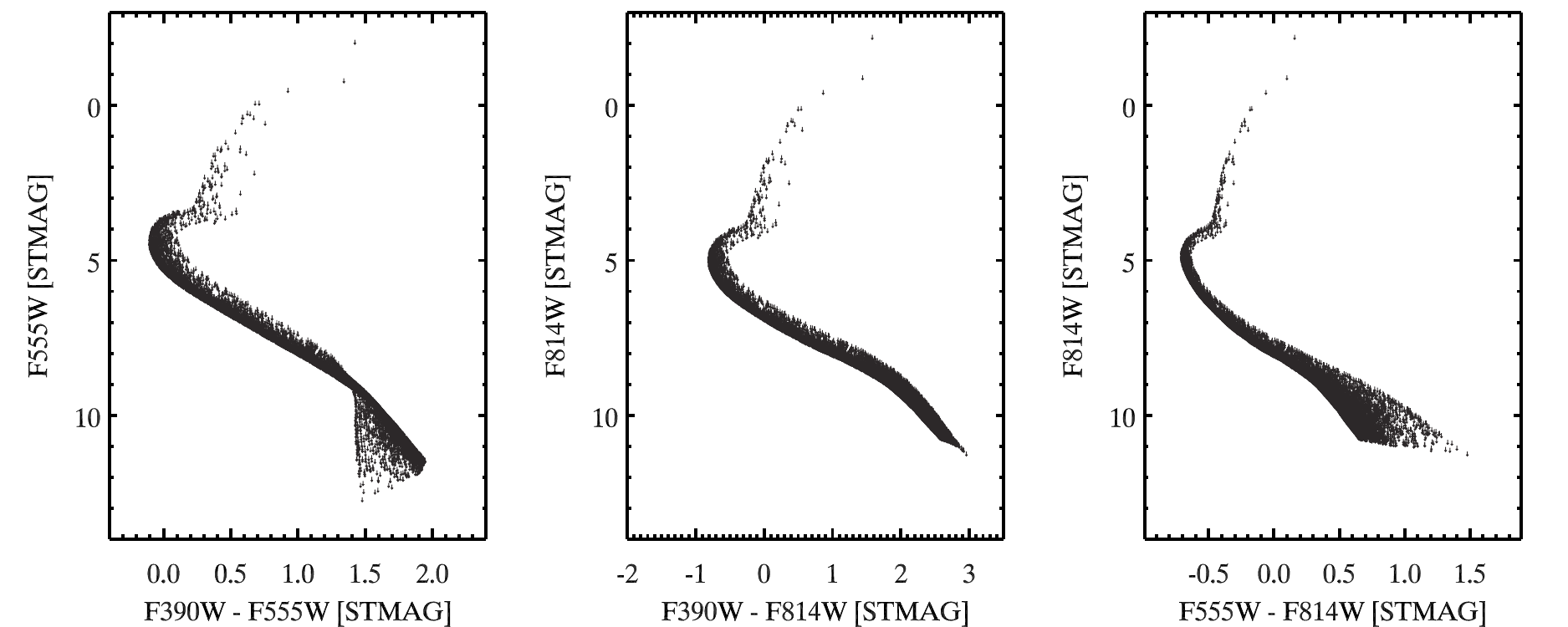}
\caption{Same as Fig.~\ref{fig:extcat0} but for the {\sc Exp} catalog \label{fig:extcat5} 
}
\end{figure*}

\begin{figure*}[!t]
\centering
\includegraphics[width=.99\textwidth]{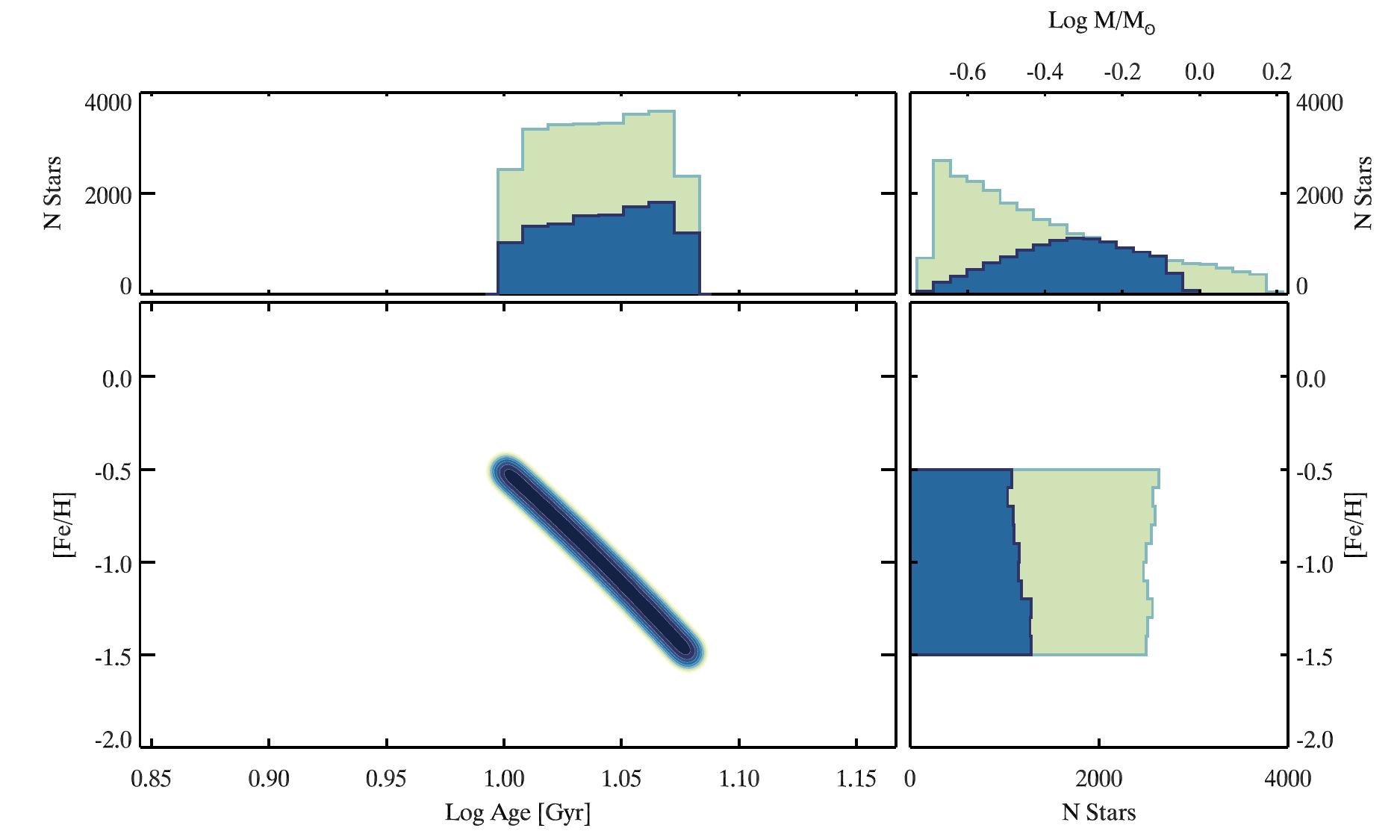}

\vspace{2pt}

\includegraphics[width=.96\textwidth]{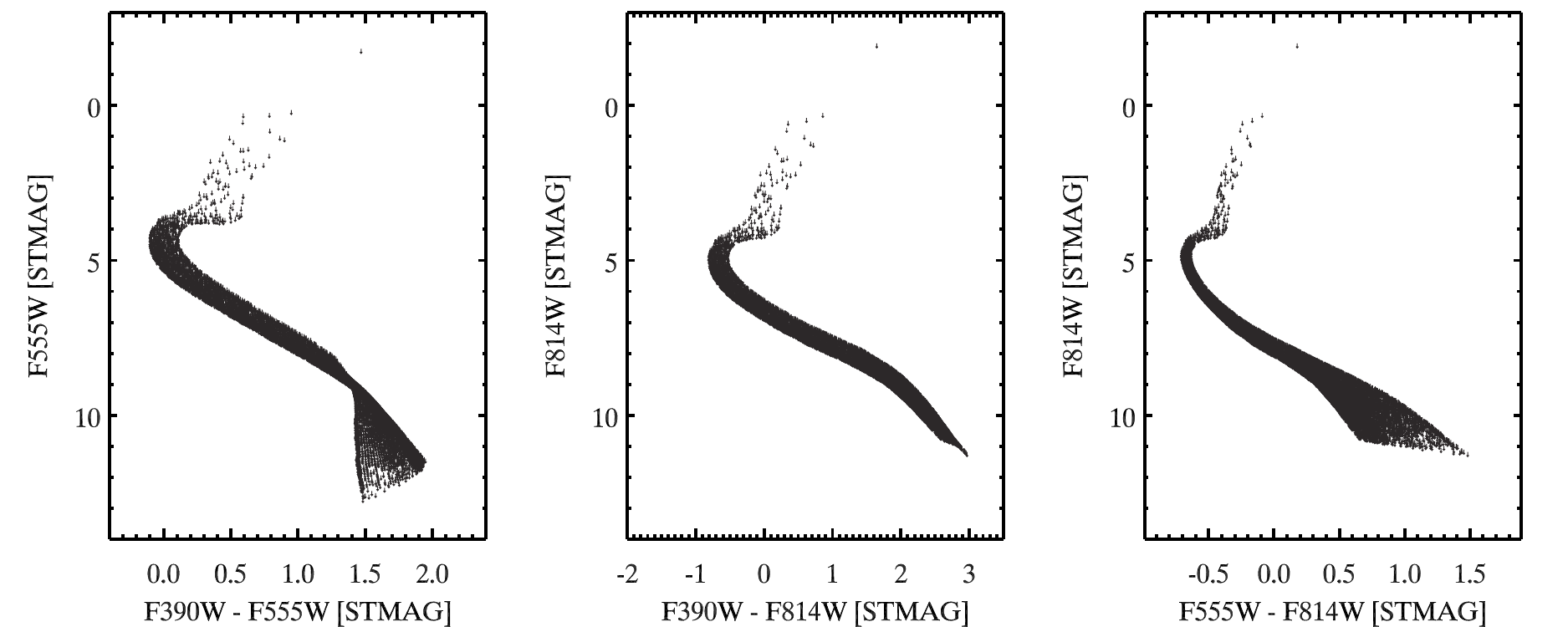}
\caption{Same as Fig.~\ref{fig:extcat0} but for the {\sc Const} catalog \label{fig:extcat6} 
}
\end{figure*}

\end{document}